\documentclass[aps,prd,twocolumn,superscriptaddress,preprintnumbers,floatfix,nofootinbib,notitlepage,showkeys,showpacs]{revtex4-1}

\usepackage[utf8]{inputenc}

\usepackage{graphicx}
\usepackage{hyperref}
\usepackage{latexsym}
\usepackage{amsmath}
\usepackage{amssymb}
\usepackage{bbm}
\usepackage{ulem}
\usepackage{pdfsync}
\usepackage{epsfig}
\usepackage{epstopdf}
\usepackage{subfigure}
\usepackage{xcolor}
\usepackage{comment}
\usepackage{slashed}
\usepackage{upgreek}

\long\def\exclude#1{}

\newcommand{\beq}{\begin{equation}}
\newcommand{\eeq}{\end{equation}}

\allowdisplaybreaks

\bibpunct{[}{]}{,}{n}{}{,}

\newcommand{\avg}[1]{\langle #1 \rangle}

\newcommand{\pdf}{\mathcal{P}}
\newcommand{\Fpdf}{\tilde{\mathcal{P}}}

\begin{document}

\title{Towards Accurate Modeling of Line-Intensity Mapping One-Point Statistics: Including Extended Profiles}
\author{Jos\'e Luis Bernal}
\affiliation{Instituto de Física de Cantabria (IFCA), CSIC-Univ. de Cantabria, Avda. de los Castros s/n, E-39005 Santander, Spain}

%==========================

\begin{abstract}
Line-intensity mapping (LIM) is quickly attracting attention as an alternative technique to probe large-scale structure and galaxy formation and evolution at high redshift. LIM one-point statistics are motivated because they provide access to the highly non-Gaussian information present in line-intensity maps and contribute to break degeneracies between cosmology and astrophysics. Now that promising surveys are underway, an accurate model for the LIM probability distribution function (PDF) is necessary to employ one-point statistics. We consider the impact of extended emission and limited experimental resolution in the LIM PDF for the first time. We find that these effects result in a lower and broader peak at low intensities and a lower tail towards high intensities. Focusing on the distribution of intensities in the observed map, we perform the first model validation of LIM one-point statistics with simulations and find good qualitative agreement. We also discuss the impact on the covariance, and demonstrate that if not accounted for, large biases in the astrophysical parameters can be expected in parameter inference. These effects are also relevant for any summary statistic estimated from the LIM PDF, and must be implemented to avoid biased results. The comparison with simulations shows, however, that there are still deviations, mostly related with the modeling of the clustering of emitters, which encourage further development of the modeling of LIM one-point statistics.

\end{abstract}

\maketitle

\section{Introduction}
\label{sec:intro}
Line-intensity mapping (LIM) is an emerging observational technique that aims to obtain three-dimensional maps of the Universe using the integrated flux of bright spectral lines over cosmological scales~\cite{Kovetz:2017agg,2012RPPh...75h6901P,Liu:2019awk, Bernal:2022jap}. Therefore, LIM probes the aggregate line emission by all sources at a given redshift. This grants access to otherwise undetectable faint populations of emitters and makes LIM the optimal tracer of the large-scale structure at high redshift in the high-noise or high-confusion regimes~\cite{Cheng:2018hox, Schaan:2021hhy}. Furthermore, dropping the requirement of resolved detection of individual emitters enables the use of low-aperture telescopes and quick scans of large portions of the sky. 

Besides tracing the underlying matter density fluctuations, line-intensity fluctuations depend on the astrophysical phenomena that trigger the line emission. Hence, combining different spectral lines, LIM also probes galaxy formation and evolution across cosmic times, complementing luminosity functions from galaxy surveys. LIM is attracting increasing attention, with numerous experiments and surveys currently observing~\cite{vanHaarlem:2013dsa, Bandura:2014gwa, DeBoer:2016tnn, MeerKLASS:2017vgf, Cleary:2021dsp, CONCERTO:2020ahk, Gebhardt:2021vfo} and many others that will see first light in the forthcoming years~\cite{CCAT-Prime:2021lly, Sun:2020mco, Switzer:2021jeg, 2020arXiv200914340V, Dore:2014cca, Koopmans:2015sua, Newburgh:2016mwi}. Preliminary detections (see e.g., Refs.~\cite{Chang:2007xk, Keating:2016pka, Yang:2019eoj, Keating:2020wlx, Wolz:2021ofa, Niemeyer:2022vrt, Cunnington:2022uzo, Niemeyer:2022arn, Paul:2023yrr}) give rise to optimism regarding the prospects for this technique and its potential to bridge between low-redshift galaxy surveys and early-Universe probes like the cosmic microwave background (see e.g., Refs.~\cite{MoradinezhadDizgah:2018lac, Bernal:2019gfq, Sato-Polito:2020cil, MoradinezhadDizgah:2021upg}). 

Most commonly, the main type of summary statistics employed to analyze line-intensity maps are 2-point statistics like the correlation function or its Fourier counterpart, the power spectrum. They benefit from some robustness against smooth, uncorrelated observational contaminants and build upon the comprehensive formalism developed for galaxy surveys. However, line-intensity fluctuations, which trace the non-linear, non-Gaussian large-scale structure and are modulated by non-trivial astrophysical processes, are very non Gaussian. Therefore, a significant part of the information contained in line-intensity maps is not captured by power spectrum measurements. Furthermore, LIM intrinsic sensitivity to both cosmology and astrophysics may hinder the interpretation of the LIM power spectra. For instance, the mean intensity and the bias relating matter and source fluctuations present a perfect degeneracy in the power spectrum at large scales. Although the degeneracy can be partially broken using smaller scales in the mildly non-linear regime~\cite{Castorina:2019zho,MoradinezhadDizgah:2021dei}, the power spectrum is only sensitive to the first two moments of the luminosity function, which result in degenerate astrophysical and cosmological parameters (see e.g., Refs.~\cite{Bernal:2019jdo, Camera:2019iwy} for a discussion on the degeneracies between astrophysics and cosmology for the power spectrum).

Accessing non-Gaussian information and breaking degeneracies between cosmology and astrophysics motivate the development of alternative summary statistics. Non-Gaussian information can be obtained using higher-order statistics like the bispectrum and trispectrum. However, one-point statistics, which depend directly on the LIM probability distribution function (PDF), hence on the full distribution of (non-Gaussian) intensity fluctuations and the whole line-luminosity function, offer preferable properties. 

The simplest one-point statistics is the actual estimator of the PDF, known as voxel intensity distribution (VID) in the context of LIM~\cite{Breysse:2015saa, Breysse:2016szq}. The VID is very complementary to the power spectrum because while the latter depends on clustering and the first two moments of the line-luminosity function, the former depends on subsequent convolutions of the luminosity function and zero-lag moments of clustering. The potential of the combination of both summary statistics has been demonstrated in the literature~\cite{COMAP:2018kem,Sato-Polito:2022fkd}. Furthermore, several studies have highlighted the sensitivity of the VID not only to astrophysical parameters but also to beyond-$\Lambda$CDM cosmologies and physics beyond the Standard Model, either directly or through reducing degeneracies when combined with the power spectrum~\cite{Bernal:2020lkd, Bernal:2021ylz, Bernal:2022wsu, Libanore:2022ntl, Adi:2023qdf}.

First instances of the LIM PDF formalism relied on small modifications to the probability-of-deflection techniques (see e.g., Refs.~\cite{1957PCPS...53..764S, 1992ApJ...396..460B, Lee:2008fm, 2009ApJ...707.1750P}): they considered the conditional probability of finding a given intensity in a voxel given number of emitters contained in it~\cite{Breysse:2015saa, Breysse:2016szq}. Ref.~\cite{Sato-Polito:2022fkd} explicitly modeled the dependence on the local emitter overdensity to derive the first analytic model of the VID supersample covariance and the VID-power spectrum covariance. Recently, Ref.~\cite{Breysse:2022alx} adapted the formalism presented in Refs.~\cite{Thiele:2018jdl, Thiele:2020rig} to compute the LIM PDF using the halo model, emphasizing the ability of the VID to break the degeneracy between the mean intensity and the emitter bias. 

Nonetheless, all previous work assumes that each emitter just contributes to the voxel in which it is contained. This in an unphysical approximation: whether it is due to extended sources, line broadening caused by the internal peculiar velocity of the gas or the limited experimental resolution, the observed emission from each emitter is redistributed to a finite volume that extends beyond the voxel in which each emitter is located. 

In this work we build upon the halo-model based formalism from Ref.~\cite{Breysse:2022alx} and model this effect in the analytic derivation of the LIM PDF. We apply the Ergodic hypothesis and obtain the PDF of the intensity in a voxel for a single emitter from the spatial profile of its observed intensity. From this starting point we can model the LIM PDF including the smearing of the observed signal due to extended sources, line broadening, and limited experimental resolution. For the first time, we show a comparison between the theoretical analytic prediction of the LIM PDF and the results for simulations to validate our model. We find qualitative good agreement, satisfactorily capturing the effects of the extended intensity profiles. Including extended observed intensity profiles significantly changes the VID prediction, and they must be taken into account to ensure unbiased parameter inference. We also find, as expected, that accounting for the fact that an emitter contributes to more than one voxel increases the correlation between intensity bins. However, other parts of the modeling, especially those related to matter clustering, must be improved to match the results from simulations. The updated VID modeling is now included in the publicly available code \textsc{lim}.\footnote{\url{https://github.com/jl-bernal/lim}}

While in this work we focus on the VID, extended emission and limited experimental resolution  must also be included in the derivation of any summary statistic that relies on the LIM PDF. Examples beyond the VID include 1-point cross correlations of different data sets, such as the conditional VID~\cite{Breysse:2019cdw} or the deconvolved density estimator~\cite{Breysse:2022fdi, COMAP:2022sdg}.

We assume the standard $\Lambda$CDM cosmology, with best-fit parameter values from the full \textit{Planck} analysis without external observations~\cite{Planck:2018vyg}. In addition, we employ the following convention for the Fourier transforms. For a $d$-dimensional Fourier-space variable $\pmb{u}$ being the conjugate of a configuration-space variable $\pmb{v}$, the direct and inverse Fourier transforms of a function $f$ and its Fourier counterpart $\tilde{f}$ are given by 
\begin{equation}
\begin{split}
    \tilde{f}(\pmb{u}) & = \int {\rm d}^d\pmb{v} \ f(\pmb{v})e^{-i\pmb{vu}}\,, \\
    f(\pmb{v}) & = \int \frac{{\rm d}^d\pmb{u}}{(2\pi)^d}\ \tilde{f}(\pmb{u})e^{i\pmb{vu}}\,, 
\end{split}
\end{equation}
where the tilde denote Fourier-space functions. Given the large dynamical range of the variables considered for Fourier transform (spanning orders of magnitude, positive and negative values), we compute the Fourier transform using Non-Uniform Fast Fourier Transform routines as implemented in \texttt{FINUFFT}\footnote{\url{https://finufft.readthedocs.io/en/latest/}}~\cite{2019SJSC...41C.479B, 2020arXiv200109405B}.

This article is structured as follows. First, we review the PDF modeling from Ref.~\cite{Breysse:2022alx} and introduce the effects of extended  intensity profiles in Sec.~\ref{sec:modeling}. We discuss limited experimental resolution and line broadening as causes of the extended profiles and their impact in the PDF prediction in Sec.~\ref{sec:eff_ext_emission}. We compare theoretical predictions with simulations in Sec.~\ref{sec:sims}. We discuss the relevance of modeling the extended emission estimating the bias introduced in parameter inference when point sources are considered in Sec.~\ref{sec:bias}. We summarize and conclude in Sec.~\ref{sec:conclusions}. 

\section{LIM 1-point PDF}
\label{sec:modeling}
In this section we review the modeling to compute the line-intensity 1-point PDF and extend it to account for extended intensity profiles. We build upon the modeling for point sources presented in Ref.~\cite{Breysse:2022alx} using the halo model, which in turn adapted the formalism introduced in Refs.~\cite{Thiele:2018jdl} and~\cite{Thiele:2020rig} for the thermal Sunyaev-Zel'dovich effect and weak lensing convergence, respectively. 

\subsection{LIM signal}
The specific line intensity per unit frequency emitted at a given position $\pmb{x}$ is given by 
\begin{equation}
    I_\nu(\mathbf{x}) = \frac{c}{4 \pi \nu  H(z)} \rho_{\rm L}(\pmb{x})\,,
\end{equation}
where $c$ is the speed of light, $\nu$ is the rest-frame frequency of the spectral line of interest, $H(z)$ is the Hubble parameter at redshift $z$ corresponding to position $\pmb{x}$, and $\rho_{\rm L}$ is the line luminosity density. Using the Rayleigh-Jeans relation, the brightness temperature can be defined from the specific intensity as
\begin{equation}
    T(\pmb{x}) = \frac{c^3 (1+z)^2}{8 \pi k_\mathrm{B} \nu^3 H(z)}\rho_{\rm L}(\pmb{x}) = X_{\rm LT}\rho_{\rm L}(\pmb{x})\,,
\end{equation}
where $k_{\rm B}$ is the Boltzmann constant, and we have defined $X_{\rm LT}$ as a redshift-dependent multiplicative factor to simplify the expressions. During this work we will use the brightness temperature as variable to describe line-intensity maps, but our approach is equally applicable to specific intensities. 

We can model the spatial distribution of the line luminosity associated to a single source located at $\pmb{x}_s$, without loss of generality, as
\begin{equation}
    \frac{{\rm d}L_s}{{\rm d}\pmb{x}}(\pmb{x}\vert \pmb{\vartheta}_s) = L_{s,{\rm tot}}(\pmb{\vartheta}_s)\varrho(\pmb{x}-\pmb{x}_s\vert \pmb{\vartheta}_s)\,,
    \label{eq:Lprofile}
\end{equation}
where $L_{s,{\rm tot}}$ is the total line luminosity of the source (i.e., integrated over its line width and point spread function) and $\varrho$ is a three-dimensional spatial emission profile with inverse-volume units that is normalized to unity: $\int{\rm d}^3\pmb{x}\varrho(\pmb{x}\vert \pmb{\vartheta}_s)=1$. Any experimental or astrophysical effect that results in a spatial smoothing of the observed signal in the final map can be embedded in $\varrho$, which is generally anisotropic. Both $L_{s,{\rm tot}}$ and $\varrho$ depend on a set of astrophysical properties of the source included in the parameter vector $\pmb{\vartheta}_s$. Then, the temperature in a specific point can be expressed as 
\begin{equation}
T(\pmb{x})=X_{\rm LT}\sum_s \frac{{\rm d}L_s}{{\rm d}\pmb{x}}(\pmb{x}\vert \pmb{\vartheta}_s)\,,
\label{eq:Tofx}
\end{equation}
where the sum is over all sources, indexed by $s$. We build upon this expression to derive the PDF for extended profiles.

\subsection{The PDF for extended profiles}
\label{sec:pdf_ext}
Since brightness temperature is an additive quantity, the PDF of the aggregate emission is the convolution of the PDF of each emitter. This calculation is much more tractable in Fourier space applying the convolution theorem. Let us define $\uptau$ as the Fourier conjugate of the brightness temperature. The Fourier transform of a given PDF $\pdf$ of the brightness temperature, 
\begin{equation}
\begin{split}
    \Fpdf(\uptau) = & \int{\rm d}T\pdf(T)e^{-iT\uptau} = \avg{e^{-iT\uptau}} = \\ = & \frac{1}{V_{\rm obs}}\int_{V_{\rm obs}} {\rm d}^3\pmb{x}e^{-iT(\pmb{x})\uptau}\,,
\end{split}
\label{eq:charac_def}
\end{equation}
is the characteristic function, and angle brackets denote average over realizations. The characteristic function  is dimensionless, and can also be obtained invoking the Ergodic hypothesis taking the average over the observed volume $V_{\rm obs}$, as done in the last equality above. 

Let us explicitly separate the halo mass $M$ from $\pmb{\vartheta}$ to ease the readability. Consider first an infinitesimal mass bin centered at $M$, for which the brightness temperature PDF is 
\begin{equation}
\begin{split}
    \pdf^{(M)}(T)&=\int{\rm d}\pmb{\vartheta}\pdf(\pmb{\vartheta}\vert M)\times \\ &\times \left[\pdf^{(M,\pmb{\vartheta})}_{N=0}\delta_{\rm D}(T)+\pdf^{(M,\pmb{\vartheta})}_{N=1}\pdf_1^{(M,\pmb{\vartheta})}(T)\right]\,,
\end{split}
\end{equation}
where we marginalize over the conditional multidimensional distribution $\pdf(\pmb{\vartheta}\vert M)$ of astrophysical properties given a halo mass (see e.g., Ref.~\cite{Zhang:2023oem}), $\pdf^{(M, \pmb{\vartheta})}_{N=x}$ is the PDF of having $x$ emitters (halos) of mass $M$ and set of properties $\pmb{\vartheta}$ contributing to a specific point, and $\pdf^{(M,\pmb{\vartheta})}_x(T)$ is the PDF of finding a temperature $T$ in a point in the space receiving contributions from $x$ emitters with such properties. If there is no emitter (i.e., $N=0$), then there is no signal and $\pdf_0(T)=\delta_{\rm D}(T)$ is the Dirac delta. For an infinitesimal mass bin $\pdf^{(M,\pmb{\vartheta})}_{N>1}=0$, hence $\pdf_{N=0}^{(M,\pmb{\vartheta})}=1-\pdf_{N=1}^{(M,\pmb{\vartheta})}$. 

The characteristic function can therefore be expressed as
\begin{equation}
    \Fpdf^{(M)}(\uptau)= 1 + \int{\rm d}\pmb{\vartheta}\pdf(\pmb{\vartheta}\vert M)\pdf^{(M,\pmb{\vartheta})}_{N=1}\left(\Fpdf^{(M,\pmb{\vartheta})}_1(\uptau)-1\right)\,.
\label{eq:PDF_probs}
\end{equation}
Although the profile $\varrho$ of the extended emission extends arbitrarily in space, in practice we can truncate it at some distance large enough that there is no sizable signal loss. Under this assumption, the signal profile covers a finite volume $V_{\rm prof}$ which can depend on $M$ and $\pmb{\vartheta}$, so that
\begin{equation}
\begin{split}
    \Fpdf^{(M,\pmb{\vartheta})}_1(\uptau)  = \int{\rm d}^3\pmb{x}{\rm d}L  \pdf(\pmb{x}\vert M,\pmb{\vartheta})\pdf(L\vert M,\pmb{\vartheta},\pmb{x})e^{-iT\uptau}\,,
    \label{eq:Fpdf_M_intro}
\end{split}
\end{equation}
where we explicitly marginalize over the position (for which $\pdf(\pmb{x}\vert M,\pmb{\vartheta})$ is uniform over $V_{\rm prof}$ and zero otherwise) and $\pdf(L\vert M,\pmb{\vartheta},\pmb{x})$ accounts for any distribution of the line luminosity given the halo mass, the astrophysical properties, and the distance to the center of the profile.\footnote{The innermost integral in Eq.~\eqref{eq:Fpdf_M_intro} is equivalent to $\pdf(T\vert M)$ in Ref.~\cite{Breysse:2022alx}.} Equation~\eqref{eq:Fpdf_M_intro} is equivalent to Eq.~\eqref{eq:charac_def} for a single source. Note that the temperature factor in the exponential depends on all the quantities that are marginalized over. 

Finally, assuming Poisson statistics, and neglecting clustering for now, 
\begin{equation}
    \pdf^{(M,\pmb{\vartheta})}_{N=1} = {\rm d}M\frac{{\rm d}n}{{\rm d}M}V_{\rm prof}(M,\pmb{\vartheta})\,,
\end{equation}
where ${\rm d}n/{\rm d}M$ is the halo mass function. For infinitesimal bins, $\pdf^{(M,\pmb{\vartheta})}_{N=1}\ll 1$, so that we can interpret Eq.~\eqref{eq:PDF_probs} as the linear expansion of the exponential
\begin{equation}
\begin{split}
     \Fpdf&^{(M)}  (\uptau)= \\ & = \exp\left\lbrace\int{\rm d}\pmb{\vartheta}\pdf(\pmb{\vartheta}\vert M)\pdf^{(M,\pmb{\vartheta})}_{N=1}\left(\Fpdf^{(M,\pmb{\vartheta})}_1(\uptau)-1\right)\right\rbrace\,.
\end{split}
\end{equation}

Now we need to extend this to all halos. Using the convolution theorem, the characteristic function for the whole population is the product of the individual characteristic functions. Thus, the characteristic function $\Fpdf^{(u)}$ for the whole population without accounting for clustering is
\begin{equation}
\begin{split}
    \Fpdf ^{(u)} & (\uptau)  = \prod\Fpdf^{(M)} (\uptau) = \\
    & = \exp\left\lbrace\int{\rm d}M{\rm d}\pmb{\vartheta}\pdf(\pmb{\vartheta}\vert M)\frac{{\rm d}n}{{\rm d}M}V_{\rm prof}(M,\pmb{\vartheta})\right. \times \\
    & \qquad\quad \times  \left(\Fpdf^{(M,\pmb{\vartheta})}_1(\uptau)-1\right)\biggl\}\,,
\end{split}
\end{equation}
where we directly have substituted the sum of exponents by the integral limit in the last equality. 

From the expression above, and assuming that the astrophysical properties are uncorrelated with clustering, it is trivial to include the effect of clustering. Clustering varies in scales much larger than the observed intensity profiles of the sources. Therefore, for a specific realization (or position), we can include the effects of clustering by adding the halo overdensity field $\delta_h\equiv (n_h-\avg{n_h})/\avg{n_h}$ to the halo mass function, where $n_h$ is the local halo number density. Therefore, we promote
\begin{equation}
    \frac{{\rm d}n}{{\rm d}M} \rightarrow \frac{{\rm d}n}{{\rm d}M}(1+\delta_h(\pmb{x},M))\,.
\end{equation}
Then, the characteristic function $\Fpdf^{(\delta)}$ accounting for clustering for a \textit{single} realization is 
\begin{equation}
    \begin{split}
    \Fpdf ^{(\delta)} (\uptau) 
    & = \exp\left\lbrace\int{\rm d}M{\rm d}\pmb{\vartheta}\pdf(\pmb{\vartheta}\vert M)\frac{{\rm d}n}{{\rm d}M}V_{\rm prof}(M,\pmb{\vartheta})\right. \times \\
    & \qquad\quad \times  \left(\Fpdf^{(M,\pmb{\vartheta})}_1(\uptau)-1\right)\delta_h\biggl\}\Fpdf^{(u)}\,.
\end{split}
\end{equation}
The global characteristic function is then the average over the realizations of $\Fpdf ^{(\delta)}$. Since $\Fpdf^{(u)}$ does not depend on the overdensities, we can take it out of the average, and we are left with the exponential of the term including $\delta_h$. We invoke the moment-generating function, which states that for a random variable $X$,
\begin{equation}
    \avg{e^X} = \exp\left\lbrace \sum_{p=1}^\infty \avg{X^p}/p! \right\rbrace\,.
    \label{eq:expansion}
\end{equation}
By definition, $\avg{\delta_h}=0$, and all terms with $p>2$ vanish for a Gaussian distribution. Although gravitational collapse induces non Gaussianities and higher-order terms should be included, we take a first approximation and truncate the moment-generating function at $p=2$. Furthermore, we relate $\delta_h$ to the underlying matter density field $\delta_m$ with a linear, mass-dependent halo bias $b_h$ so that $\delta_h^{(M)}=b_h(M)\delta_m$ and the second cumulant of the halo distribution is
\begin{equation}
    \avg{\delta_h^{(M)}(\pmb{x})\delta_h^{(M')}(\pmb{x})}  \equiv b_h(M)b_h(M')\sigma^2\,,
\end{equation}
where $\sigma^2$ is the zero-lag variance of the matter distribution. 

Finally, the overall characteristic function is given by
\begin{equation}
    \begin{split}
    \Fpdf(\uptau) 
    & = \exp\left\lbrace\left[\int{\rm d}M{\rm d}\pmb{\vartheta}\pdf(\pmb{\vartheta}\vert M)\frac{{\rm d}n}{{\rm d}M}V_{\rm prof}(M,\pmb{\vartheta})\right.\right. \times \\
    & \qquad\quad \times  \left(\Fpdf^{(M,\pmb{\vartheta})}_1(\uptau)-1\right)b_h(M)\biggl]^2\frac{\sigma^2}{2}\biggl\}\Fpdf^{(u)}\,.
\end{split}
\label{eq:Fpdf_pointsource}
\end{equation}

We can obtain the PDF of the brightness temperature in a point by computing the inverse Fourier transform of the characteristic function above. Sometimes it can be useful to consider the PDF of brightness temperature fluctuations $\Delta T\equiv T-\bar{T}$ as a crude approximation for foreground subtraction, where $\bar{T}$ is the mean brightness temperature. This can be obtained multiplying the characteristic function by $e^{i\bar{T}\uptau}$.

\subsection{Voxelized volume and practical considerations}
The derivation in the subsection above returns the PDF of the brightness temperature in a specific point. In practice, however, we measure the brightness temperature from observations in a discretized map, a data cube in which each cell or voxel (three-dimensional pixels) corresponds to a comoving volume $V_{\rm vox}$. Discretizing the map is relevant at three stages of the derivation above: the volume of the profiles, the characteristic function of a single specific emitter, and the cumulants of the matter overdensity field.

After discretizing the observed volume, the measured temperature $T_i$ in a voxel centered at $\pmb{x}_i$ is
\begin{equation}
    T_i = \frac{X_{\rm LT}}{V_{\rm vox}}\sum_s \int_{V_{{\rm vox},i}}{\rm d}^3\pmb{x}\frac{{\rm d}L_s}{{\rm d}\pmb{x}}(\pmb{x}\vert \pmb{\vartheta}_s)\,,
    \label{eq:Tvox}
\end{equation}
where the sum is over all sources indexed by $s$. 

For isotropic intensity profiles and spherical voxels, the right hand side of Eq.~\eqref{eq:Fpdf_M_intro} can be easily evaluated inverting the relationship between temperature and position for a given emitter (see appendix B in Ref.~\cite{Thiele:2020rig}). However, this is not possible in a more general setup. In particular, the observed signal profiles in LIM surveys are anisotropic due to different angular and spectral resolutions, to line broadening only affecting the profile in the direction along the line of sight, etc. Moreover, although the voxelization can be arbitrary it is more optimal to follow the experimental resolutions, which do not need to correspond to similar distances in the direction along and transverse to the line of sight. 

Instead, we take a different approach and explicitly compute the spatial integral in Eq.~\eqref{eq:Fpdf_M_intro}, adapted for a voxelized space. First, we consider a three-dimensional space and locate the emitter of interest in its center. We grid the space with cells with the same comoving size as the voxels and compute the total luminosity on each voxel with the integral of Eq.~\eqref{eq:Tvox}. We truncate the emission profile at a minimum relative luminosity $L_{\rm rel}^{\rm min}$ with respect to the voxel with maximum luminosity (e.g., keeping only the $N_{\rm vox}^{\rm prof}$ voxels with $L_{\rm vox}/{\rm max}(L_{\rm vox})\geq L_{\rm rel}^{\rm min}$). We normalize the luminosity on each voxel so that their sum is again $L_{s,{\rm tot}}$, and consider $V_{\rm prof}=N_{\rm vox}^{\rm prof}V_{\rm vox}$. 

Then, we take advantage of the scaling property of the Fourier Transform. Let us consider some scale temperature $T_0$ so that any $T=CT_0$. We can assume any PDF $\pdf_{T_0}$ for $T_0$ to account for any scatter in the conditional relations considered in the previous subsections. We assume a lognormal distribution
\begin{equation}
    \pdf_{T_0}(T;T_0) = \frac{\exp\left\lbrace \frac{-\log_{10}\left(\frac{T}{T_0}\right) - \frac{\sigma_{\rm scat}^2\log(10)}{2}}{2\sigma_{\rm scat}^2}\right\rbrace}{\sqrt{2\pi}T\sigma_{\rm scat}\log(10)}
    \label{eq:lognormal}
\end{equation}
with mean $T_0$ and scatter $\sigma_{\rm scat}$ in dex, so that for any $T'\equiv CT_0$ we have $\pdf_{T'}(T;CT_0)=\pdf_{T_0}(T/C;T_0)/C$, which fulfills $\Fpdf_{T'}(\uptau;CT_0) = \Fpdf_{T_0}(C\uptau;T_0)$. Grouping all $N_{\rm vox}^{(i)}$ voxels with the same temperature $T_i$, the characteristic function for a halo with mass $M$ and properties $\pmb{\vartheta}$ is
\begin{equation}
    \Fpdf_1^{(M,\pmb{\vartheta})}(\uptau) = \sum_i \frac{N_{\rm vox}^{(i)}}{N_{\rm vox}^{\rm prof}}\Fpdf_{T_0}(T_i\uptau/T_0;T_0)\,,
\end{equation}
which corresponds to the weighted average of the characteristic function for each position in the spatial profile, i.e., a discretized version of Eq.~\eqref{eq:Fpdf_M_intro}.  

Finally, the discretization also affects the impact of clustering in the PDF. Voxels are the basic unit of information we have access to, hence halo overdensities are smoothed over scales of the size of the voxel. This can be modeled by convolving the overdensity field with the window function $W_{\rm vox}$ of the voxel, usually a normalized top-hat function with the extent of the voxel, leaving
\begin{equation}
    \delta_h^{\rm v}(\pmb{x}) = \int {\rm d}^3\pmb{x}'W_{\rm vox}(\pmb{x}-\pmb{x}')\delta_h(\pmb{x}'),
\end{equation}
where $\int{\rm d}^3\pmb{x}W_{\rm vox}=1$. We substitute $\delta_h$ by $\delta_h^{\rm v}$, and similarly for $\delta_m$, in all their instances in the previous subsection. This change can be summarized with a slight modification in Eq.~\eqref{eq:Fpdf_pointsource}, where the cumulant of matter perturbations must be
\begin{equation}
    \sigma_{\rm vox}^2 = \int\frac{{\rm d}^3\pmb{k}}{(2\pi)^3} \tilde{W}^2_{\rm vox}(\pmb{k})P_m^{(s)}(k)\,,
    \label{eq:sigma_vox}
\end{equation}
$P_m^{(s)}$ is the nonlinear matter power spectrum in redshift space and $\tilde{W}_{\rm vox}$ is the Fourier transform of the voxel window function. We estimate the nonlinear power spectrum as modeled by \textsc{HMcode}~\cite{Mead:2020vgs}. In this work, we model redshift-space distortions with the Kaiser factor and a phenomenological Lorentzian suppression to model the fingers of God (following e.g., Ref.~\cite{Bernal:2019jdo}) with characteristic scale 
\begin{equation}
    \sigma_{\rm FoG} = \frac{4\pi}{3}\int \frac{{\rm d}k}{(2\pi)^3}P_m^{{\rm lin}}(k)\,,
    \label{eq:sigma_FoG}
\end{equation}
where $P_m^{\rm lin}$ is the linear matter power spectrum in real space.

Still, the PDF is a continuous quantity which cannot be directly measured from the observations. Instead, the VID $\mathcal{B}$, a histogram of the measured brightness temperature in each voxel normalized by the total number of voxels, can be used as an estimator of the PDF. The relation between the VID and the PDF of the signal (i.e., in the absence of experimental thermal noise) is\footnote{Previous studies do not normalize the VID by the number of voxels observed, and therefore include a factor $N_{\rm vox}$ multiplying the integral in Eq.~\eqref{eq:VID}. We prefer to normalize the VID to deal with an intensive quantitative, the value of which does not depend on the size of the survey. This approach allows for a more intuitive understanding of the VID values, as well as easier comparisons between experiments.}
\begin{equation}
    \mathcal{B}^{\rm (s)}(\Delta T_i) = \int_{\Delta T_i}{\rm d} \Delta T \pdf(\Delta T)\,,
    \label{eq:VID}
\end{equation}
where the integral is limited to the temperature interval centered on $\Delta T_i$. Nonetheless, there is no perfect experiment without noise, and the total VID $\mathcal{B}$ is connected to the total PDF $\pdf_{\rm tot}(\Delta T) = (\pdf*\pdf_{\rm noise})(\Delta T)$  through Eq.~\eqref{eq:VID}, where $\pdf_{\rm noise}$ is the instrumental noise PDF and `*' denotes the convolution operator. 

\section{Effects from extended emission}
\label{sec:eff_ext_emission}
The derivation above is general enough to account for any three-dimensional signal profile and any source that can be embedded in the halo model formalism. Suppose a series of effects that cause extended signal which independently correspond to three-dimensional profiles (or window functions) $W_i$, normalized to unity. Then, the final extended observed profile is given by the convolution of all of them:
\begin{equation}
    \varrho(\pmb{x}) = \left(W_1*W_2*\dots *W_n\right)(\pmb{x})\,.
    \label{eq:general_prof}
\end{equation}

As an example, we limit our study to emission coming from halos and to two of the most prevalent causes of extended profiles: experimental resolution and line broadening. The former relates to the fact that any experiment has limited resolution, which prevents to access small scale information and smears the observed maps; the latter is a physical effect caused by the Doppler broadening of the emission line due to the peculiar velocities of the gas where the emission originates.\footnote{In the case of the Lyman-$\alpha$ line, radiative transfer also affects the observed signal profile; we leave the study of this effect for future work.} 

\subsection{Limited resolution and line broadening}
The small-scale information in line-intensity maps is limited by the angular and spectral resolutions of the experiment. The exact characterization of the beam profiles and the line-spread function depend on each experiment. Here, we will consider a Gaussian beam with full-width half maximum $\theta_{\rm FWHM}$ and a Gaussian line-spread function with standard deviation given by the channel width $\delta\nu$ of the experiment. The standard deviation of the Gaussian window functions for these resolutions correspond to comoving distance scales transverse to and along the line of sight given by
\begin{equation}
\begin{split}
    \sigma_\perp  = D_{\rm M}(z)\frac{\theta_{\rm FWHM}}{\sqrt{8\log 2}}\,, \qquad
    \sigma_\parallel = \frac{c\delta\nu(1+z)}{H(z)\nu_{\rm obs}}\,,
\end{split}
    \label{eq:sigma_perp_parallel}
\end{equation}
respectively, where $D_{\rm M}$ is the comoving angular diameter distance, and $\nu_{\rm obs}$ is the observed frequency. The window function $W_{\rm res}= W_{\rm res}^\perp W_{\rm res}^\parallel$ for the resolution is composed of 
\begin{equation}
\begin{split}
     W_{\rm res}^\perp(\pmb{x}_\perp) & =\frac{\exp\left\lbrace -(x_{\perp,1}^2+x_{\perp,2}^2)/2\sigma_\perp^2\right\rbrace}{\sqrt{2\pi}\sigma_\perp}\,,\\ 
     W_{\rm res}^\parallel(x_\parallel) & = \frac{\exp\left\lbrace -x_\parallel^2/2\sigma_\parallel^2\right\rbrace}{\sqrt{2\pi}\sigma_\parallel}\,,
\end{split} 
\label{eq:Wk_res}
 \end{equation} 
where the subscript `1' or `2' denote the directions of the axis transverse to the line of sight. 

We assume that the bulk of the line broadening is due to the rotation of the gas in the halo, which in turn only depends on the halo mass. Depending on the line of interest, this may not be the case, especially if the line is not emitted by hot gas but by colder gas in the galactic disk.  In the latter case, the width of the line also depends on the disk mass. In this work, we limit our focus to the halo-mass dependent model proposed in Ref.~\cite{COMAP:2021rny} and leave the study of further dependencies to future work.

Let us consider that the broadening follows a Gaussian profile with full-width half maximum determined by the rotation velocity $v(M)$ in units of physical velocity. The standard deviation of the Gaussian profile in comoving space is
\begin{equation}
    \sigma_v(M) = \frac{v(M)(1+z)}{2\sqrt{2\log(2)}H(z)}\,.
    \label{eq:sigmav}
\end{equation}
Line broadening effectively reduces the map resolution along the line of sight. Therefore, the window function capturing its effect is the same $W_{\rm res}^\parallel$ from Eq.~\eqref{eq:Wk_res}, but using $\sigma_v$ instead of $\sigma_\parallel$. 

However, a strictly Gaussian profile would consider that all galaxies have the same angle of inclination with respect to the line of sight. A much more reasonable assumption involves randomly oriented emitters, as considered in Ref.~\cite{COMAP:2021rny}. The actual standard deviation for a galaxy with an inclination angle $\varphi$ is $2\sigma_v\sin\varphi/\sqrt{3}$, assuming that the previously calculated $\sigma_v$ corresponds to galaxies with median inclination over random inclinations. The average profile can be obtained marginalizing over the inclination of the galaxy, which follows a uniform distribution on $\cos\varphi \equiv \mu_\varphi$:
\begin{equation}
\begin{split}
    W_{\rm broad}& (x_\parallel,M) = \\ 
    = & \int_{-1}^{1}{\rm d}\mu_\varphi\frac{\exp\left\lbrace -3x_\parallel^2/8\sigma_v^2(M)(1-\mu_\varphi^2)\right\rbrace}{4\sigma_v(M)\sqrt{2\pi(1-\mu_\varphi^2)/3}} = \\
    = & \frac{\sqrt{3\pi/2}}{4\sigma_v}{\rm Erfc}\left(\sqrt{3x_\parallel^2/8\sigma_v^2}\right)\,,
\end{split}
\label{eq:Wbroad}
\end{equation}
where Erfc is the complementary error function. 

If we assume that these are the only contributions to extended signal, applying Eq.~\eqref{eq:general_prof} we find
\begin{equation}
\begin{split}
    \varrho(\pmb{x},M) & = W_{\rm res}(\pmb{x})*W_{\rm broad}(x_\parallel,M) = \\ 
    & = W_{\rm res}^\perp(\pmb{x}_\perp)\left(W_{\rm res}^\parallel(x_\parallel)*W_{\rm broad}(x_\parallel,M)\right)\,.
\end{split}
\end{equation}
If we were to ignore the effect of random inclinations for the line broadening, then $\varrho^\parallel$ would be a Gaussian with standard deviation $\sqrt{\sigma^2_\parallel+\sigma_v^2(M)}$. However, we have not found an analytic expression for the convolution of a Gaussian and Eq.~\eqref{eq:Wbroad}. Numerically, it can be quickly computed using the convolution theorem. For reference, the Fourier transform of Eq.~\eqref{eq:Wbroad} is
\begin{equation}
    \tilde{W}_{\rm broad}(k_\parallel) = \left(\sqrt{\frac{2}{3}}k_\parallel\sigma_v\right)^{-1}F\left(\sqrt{\frac{2}{3}}k_\parallel\sigma_v\right)\,,
\end{equation}
where $F$ is the Dawson integral.

\subsection{Experimental case and astrophysical model}
\label{sec:exp_astro}
By design, the VID depends not only on the intrinsic signal but also on the experimental setup of the LIM survey, e.g., through the noise, the resolution, the voxel volume, etc. For instance, if the spectral resolution is bad enough, the effect of line broadening will be negligible. This is why we need to focus on a specific case and generalize the qualitative effect of extended signal in the VID. To augment the effect of line broadening, we choose good experimental resolution. 

We consider the  spectral line CO(1-0)  observed at $\nu_{\rm obs}=28.8$ GHz ($z=3$), with $\theta_{\rm FWHM}=3$ arcmin and $\delta\nu = 15$ MHz, and a noise-per-voxel standard deviation of $\sigma_{\rm N}=2.5\,\mu{\rm K}$. This ensures some effect of the line broadening and that the VID is only dominated by noise at small brightness temperatures. Since we are interested in the VID prediction, we assume a good understanding of the foregrounds, calibration and sky subtraction. 

The optimal pixel size for a projected angular map, balancing the minimization of the correlation between different intensity bins and the number of pixels to reduce the statistical uncertainties corresponds to $\theta_{\rm FWHM}$~\cite{Vernstrom:2013vva}. We find the same to be true for the direction along the line of sight. Therefore, we choose our voxel size in the direction transverse to and along the line of sight as the correspondent to $\theta_{\rm FWHM}$ and $\delta\nu^{\rm FWHM}\equiv \delta\nu\sqrt{8\log 2}$, respectively. 

We model the mean relationship between the total line luminosity $L_{s,{\rm tot}}$ and the halo mass using the fiducial COMAP model~\cite{COMAP:2021lae},
\begin{equation}
    \frac{L_{\rm CO}}{L_{\odot}}(M) = 4.9\times 10^{-5}\frac{C}{\left(M/M_\star\right)^A+\left(M/M_\star\right)^B}\,,
    \label{eq:LofM}
\end{equation}
with $A=-2.85$, $B=-0.42$, $C=10^{10.63}$, and $M_\star=10^{12.3}\,M_\odot$, obtained from a fit to results from Universe Machine~\cite{Behroozi:2019kql}, COLDz~\cite{Riechers:2018zjg} and COPSS~\cite{Keating:2016pka}. We apply a mean-preserving scatter $\sigma_{\rm scat}=0.42$ dex in the expression above, introduced in practice in Eq.~\eqref{eq:lognormal}. We assume the halo mass function and mass-dependent linear halo bias from Ref.~\cite{Tinker:2008ff} and Ref.~\cite{Tinker:2010my}, respectively.

Probably the best estimation of the gas dynamics in a halo comes from the maximum circular velocity calculated by a halo finder in a simulation. However, there is no given relationship between this velocity and the halo mass. We choose to use the virial velocity as a good approximation, given by 
\begin{equation}
    v(M) = 50\,{\rm km/s} \left(\frac{M}{10^{10}\,M_\odot}\right)^{1/3}\,,
    \label{eq:vofM}
\end{equation}
to compute the line width with Eq.~\eqref{eq:sigmav}. We refer the interested reader to Ref.~\cite{COMAP:2021rny} for a comprehensive discussion on line broadening, covering the estimation of $v(M)$ from observations, and the modeling of the line broadening in simulations and in the prediction of the power spectrum multipoles.

\subsection{Changes in the PDF}
Intituively, we can expect that extended signal profiles smooths the PDF: as each source contributes to more than one voxel, there are less very faint or very bright voxels, since the signal gets redistributed and homogenized. This scenario corresponds to a PDF with a smoother, smaller peak at low brightness temperatures and a smaller tail extending to high brightness temperatures. In this section we check this intuition by studying the effects of experimental resolution and line broadening in the PDF. 

\begin{figure}[t]
\centering
\includegraphics[width=\columnwidth]{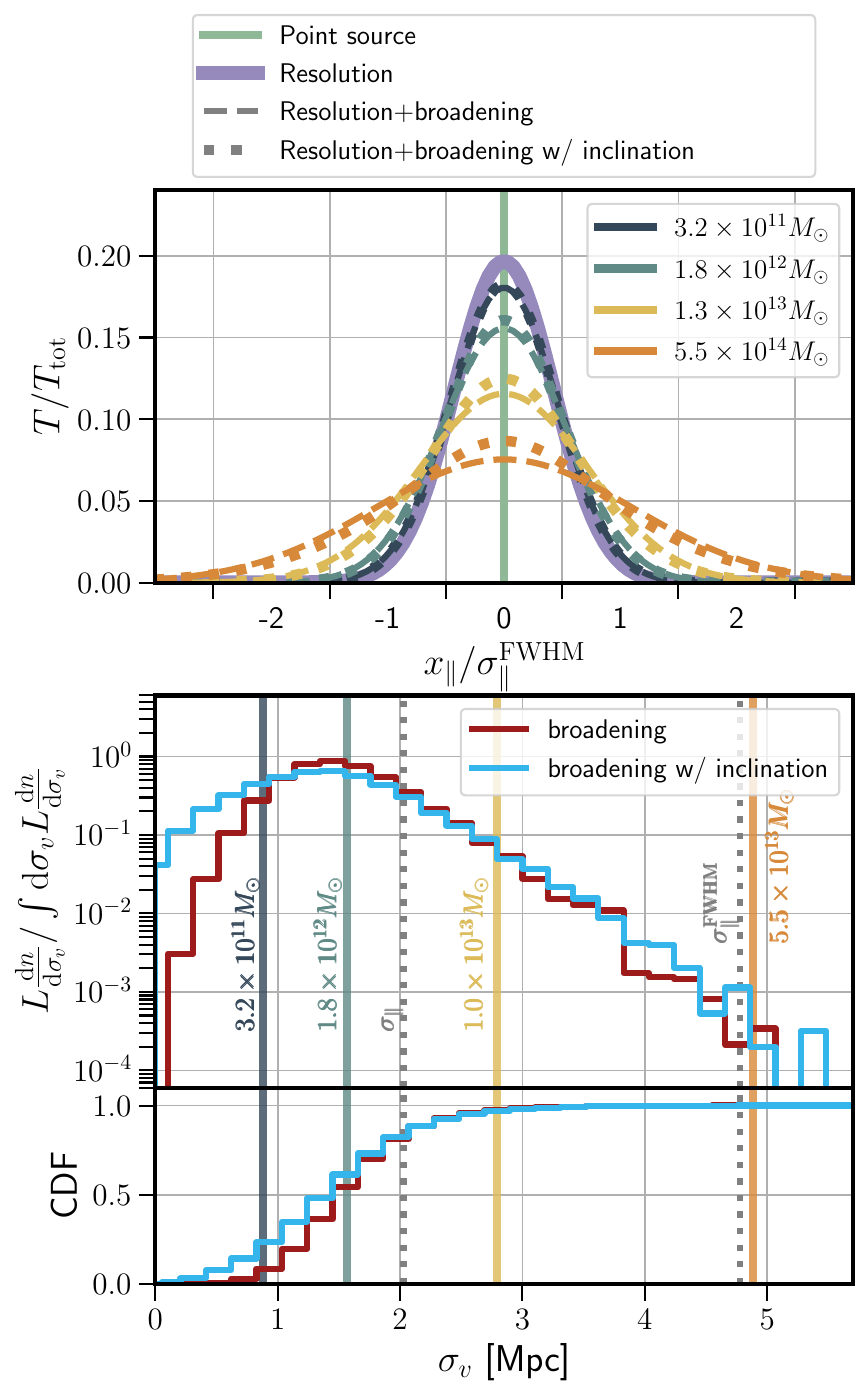}
\caption{Top panel:  extended profiles in the direction along the line of sight for our example under different assumptions: we show a point source  in light green, the effects of the limited resolution in purple, and adding line broadening with and without random inclination (dotted and dashed lines, respectively) with different colors for different halo masses. The grid in the horizontal axis corresponds to the voxel division. Bottom panel: Luminosity-weighted distribution (top) and cumulative distribution (bottom) of the line broadening with and without random inclinations (blue and red, respectively). We also show the line broadening $\sigma_v$ for edge-on emitters for the same masses considered in the top panel, the standard deviation $\sigma_\parallel$ of the line-spread function and the voxel size $\sigma_\parallel^{\rm FWHM}$.}
\label{fig:profile}
\end{figure}

First, let us study the spatial distribution of the signal along the line of sight. 
We compare the profile along the line of sight under different assumptions in the top panel of Fig.~\ref{fig:profile}. We include results for a point source, accounting only for the limited resolution of the experiment, and including line broadening with and without marginalizing over the inclination of the emitter. Limited resolution spreads the signal over three voxels,\footnote{Note that for the more standard voxel size corresponding to $\delta\nu$ instead of $\delta_\nu^{\rm FWHM}$ the signal extends over 5-7 voxels.}  but line broadening results in an even more extended profile for the most massive halos. Marginalizing over random inclinations results in a slightly more peaked profile at the center, but very similar tails. 

For our example, only halos with masses $\gtrsim 4\times 10^{12}\,M_{\odot}$ have a line broadening larger than the line-spread function. However, as can be seen in the Figure, accounting for the two effects together already results in wider profiles for lighter halos. This, of course, depends on the voxel size and the specific $v(M)$ relation considered. We can quantify the relative contribution to the intensity map from the emitters that extend beyond the resolution limits with the luminosity-weighted distribution of $\sigma_{v}$ (bottom panel of Fig.~\ref{fig:profile}). The contribution of halos with emission lines broader than the line-spread function is small, because although they are the brightest emitters, their abundance is very small. We therefore expect a small effect in the VID with respect to considering only the experimental resolution. Finally, the random inclination of the emitters broadens the distribution of $\sigma_v$, but have a similar relative contribution to the line-intensity map (as seen with the cumulative distribution function). 

\begin{figure}[t]
\centering
\includegraphics[width=\columnwidth]{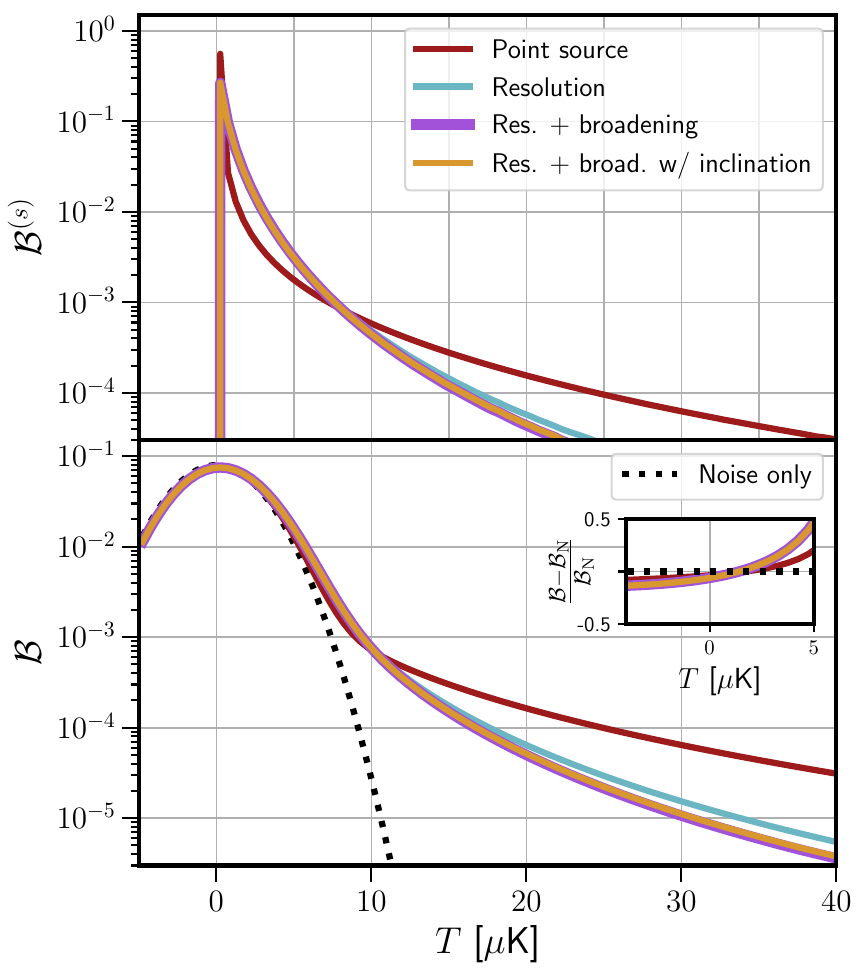}
\caption{VID of the signal (top) and including the contribution from the instrumental noise (bottom) for our example. We show the VID for point sources in red, including the effects from limited resolution in blue, adding line broadening for edge-on galaxies  in magenta, and marginalizing also over their inclination in orange. The inset zooms in on the deviation over the noise-alone VID (black-dotted).}
\label{fig:VID}
\end{figure}

We show the effects of the extended profiles in the PDF in Fig.~\ref{fig:VID}, considering the same cases than in Fig.~\ref{fig:profile}. We show the VID of the signal alone and including the noise, using 90 linearly spaced bins of 0.5 $\mu$K width in the interval $T \in [-5,40]\,\mu{\rm K}$, as well as the VID corresponding to only noise. The effects induce a very significant change in the VID, following the intuition introduced at the beginning of the section. The extended emission narrows and smooths the VID, resulting in a lower tail towards higher brightness temperatures and a broader, lower peak at low brightness temperatures. Adding fractional contributions of several emitters to voxels which otherwise would be empty results in less very faint voxels. At the same time, distributing the signal of very bright voxels over their adjacent voxels imply that they are not that bright any more. 

As one could expect from the results shown in Fig.~\ref{fig:profile}, the impact of including the line broadening on top of the experimental resolution is small and limited only to the height of the tail at high temperatures. Marginalizing over the galaxy inclination has a negligible effect for this example, but it may have some impact for a different setup. Note that, although the VID at high brightness temperatures is significantly smaller after considering extended emission, we can expect better sensitivity to the signal of interest, since there is a larger deviation from the noise-only signal at low brightness temperatures, where the statistical uncertainties are smaller.  Figure~\ref{fig:profile} is a clear example that extended profiles  must be taken into account when computing the LIM PDF.

The redistribution of signal to different voxels due to the extended profiles has another consequence for the interpretation of the VID. While the identification of bright voxels as those hosting more massive halos or brighter emitters is already non trivial in the case of point sources (due to the unknown number of emitters in each voxel), %for point sources the identification of bright voxels with more massive halos or brighter emitters is already non trivial due to the unknown number of emitters in each voxel, 
the redistribution of signal further smears this connection. This results in a very broad distribution of halo masses without large differences between bright or faint voxels, as found in Ref.~\cite{Sato-Polito:2022wiq}.

\begin{figure}[t]
\centering
\includegraphics[width=\columnwidth]{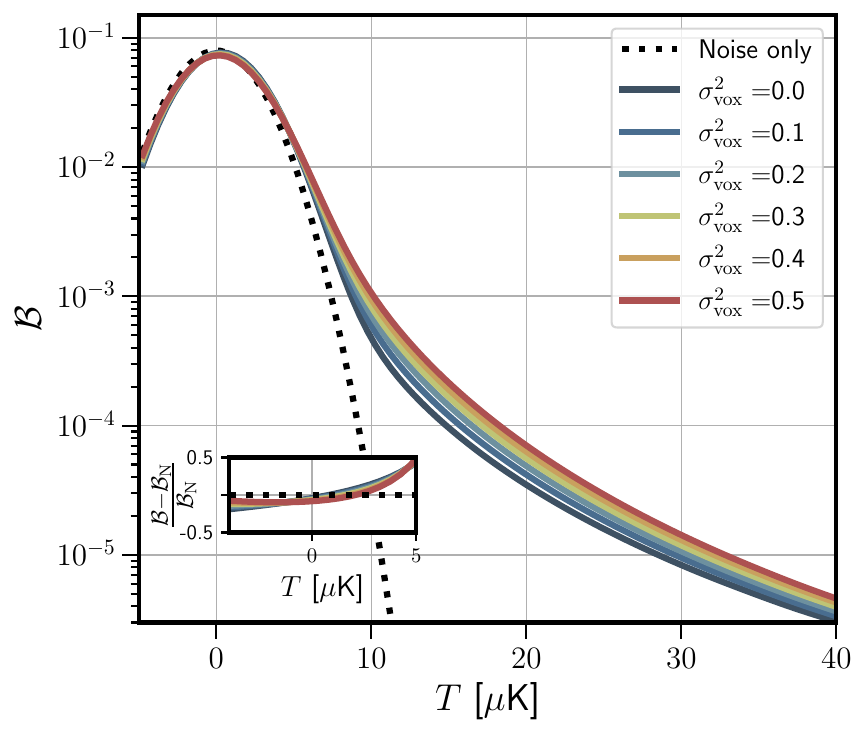}
\caption{VID including the contribution from the instrumental noise for our example and accounting for experimental resolution and the line broadening for randomly oriented galaxies. We show the prediction for several values of $\sigma_{\rm vox}^2$ with different colors, as indicated in the legend, and the noise-only VID with a black-dotted line. The inset in the bottom panel zooms in on the deviation over the noise-alone VID.}
\label{fig:clustering}
\end{figure}

We show the dependence of the VID on clustering in Fig.~\ref{fig:clustering}. Higher matter clustering implies a higher clustering of galaxies and subsequently a larger variance in the brightness temperature fluctuations. Therefore, it increases the high-temperature tail of the VID, compensating at temperatures slightly above zero, as shown in the figure and in the inset, respectively. Since this effect can be degenerate with the $L(M)$ relation, analyses must marginalize over $\sigma_{\rm vox}^2$ to avoid biased results. 

Perhaps surprisingly, line broadening has a smaller impact on the VID than in the power spectrum multipoles (see Ref.~\cite{COMAP:2021rny}). This is because line broadening is more severe for brighter, more massive, but much less abundant halos. While they dominate the contributions to the shot noise of the power spectrum (which depends on the second moment of the luminosity function) and the linear power spectrum (since their fluctuations are more biased), they almost do not contribute to the overall luminosity (see Fig.~\ref{fig:profile}). On top of that, the impact of the bias is smaller and very distributed over all masses in the VID (see Eq.~\eqref{eq:Fpdf_pointsource}). Furthermore, the quadrupole of the power spectrum traces the anisotropy of the clustering, hence it can be very affected by line broadening, while the VID is not sensitive to anisotropies. 

\section{Validation with simulations}
\label{sec:sims}
In this section we compare our predictions with simulations to test the accuracy of the modeling presented in Sec.~\ref{sec:modeling}. We assume the astrophysical model and experimental setup described in Sec.~\ref{sec:exp_astro}. 

We employ the publicly available code \textsc{Simple}\footnote{\url{https://github.com/mlujnie/simple}}~\cite{Niemeyer:2023yeu}, based on lognormal galaxy distribution simulations\footnote{\url{https://bitbucket.org/komatsu5147/lognormal_galaxies/src/master/}}~\cite{Agrawal:2017khv}, to compute the LIM lognormal simulations. We consider the luminosity-weighted average linear halo bias ($b=4.07$) corresponding to our model. We generate line-intensity maps for a volume of 0.158 $({\rm Gpc}/h)^3$ at a single redshift $z=3$, which corresponds to 150$^3$ cuboid-like voxels, defined as the full-width half maximum of the telescope beam and channel width. We run the simulations with a mesh three times finer (e.g., 450$^3$ cells) in order to accurately capture the smoothing and the extended emission. We assign the intensity emitted by each galaxy to the mesh using the `nearest grid point' routine, apply as many filters as needed depending on the case to smooth the simulated line-intensity map, and then downsample to the voxel-size resolution using the same routine.\footnote{While `nearest grid point' introduces aliases and ringing for power spectrum measurements (see Ref.~\cite{Sefusatti:2015aex} for a thorough study), we have tested that higher-order mass-assignment kernels such as `clouds in cell' introduce spurious effects in the VID, even after compensating their effect on the map by deconvoluting the kernel. The finer grid also prevents from resolution-dependent artifacts related with the Fourier transform in the final map. } 

For each realization, we consider four different cases: no smoothing applied (i.e., point sources); modeling the experimental resolution with an anisotropic Gaussian filter with standard deviations $\sigma_\parallel = 1.37\, {\rm Mpc}/h$ and $\sigma_\perp = 1.62\,{\rm Mpc}/h$ along and transverse to the line of sight, respectively; and including on top of it the line broadening with and without accounting for random inclinations. 

We follow Ref.~\cite{COMAP:2021rny} to implement line broadening in the simulations. We first compute the line width $\sigma_v$ for each galaxy. To do so, we modify $\textsc{Simple}$ to first assign a mass to each galaxy following the halo mass function. We can compute the line luminosity and line width associated with each mass using  Eq.~\eqref{eq:LofM} with a mean-preserving logarithmic scatter $\sigma_{\rm scat}$, and  Eqs.~\eqref{eq:vofM} and~\eqref{eq:sigmav}, respectively. Random galaxy inclinations are accounted for multiplying the edge-on $\sigma_v$ by a factor $2\sqrt{(1-\mu_\varphi^2)/3}$, where $\mu_\varphi$ is sampled from a uniform distribution within $[-1,1]$. We bin galaxies in the simulation by $\sigma_v$ and compute an intensity-weighted average $\bar{\sigma}_v^i$ for each bin. Afterwards, we add the intensity from the galaxies within each bin to empty meshes and apply a Gaussian filter with $\sigma_\perp = 0$ and $\sigma_\parallel = \bar{\sigma}_v^i$. We add all meshes and apply the resolution filter described above. 

In all cases, we remove the mean from the maps, measure the VID for $\Delta T$, and compute their average
\begin{equation}
    \bar{\mathcal{B}}_a = \frac{1}{N_{\rm r}}\sum_i^{N_{\rm r}} \mathcal{B}^{(i)}_a
\end{equation}
and covariance
\begin{equation}
\begin{split}
    \mathcal{C}_{ab} = \frac{1}{N_{\rm r}}\sum_i^{N_{\rm r}} \left\lbrace\left(\mathcal{B}^{(i)}_a-\bar{\mathcal{B}}_a\right) \left(\mathcal{B}^{(i)}_b-\bar{\mathcal{B}}_b\right)\right\rbrace
\end{split}
\label{eq:covariance}
\end{equation}
for the $\Delta T_a$ and $\Delta T_b$ bins from $N_{\rm r}=1008$ independent realizations. Note that we have defined $\mathcal{B}_a\equiv \mathcal{B}(\Delta T_a)$.

\begin{figure}[t]
\centering
\includegraphics[width=\columnwidth]{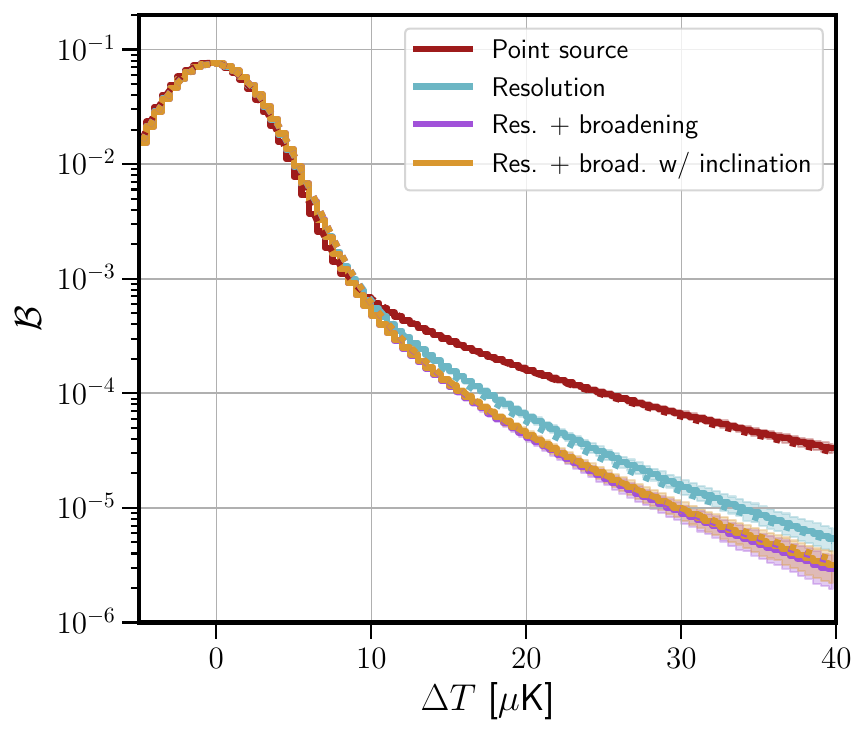}
\caption{Mean measured VID from 1008 independent lognormal realizations including instrumental noise (solid step lines) and the prediction using the formalism described in this work (dotted lines). Shaded regions correspond to the square root of the diagonal of the covariance. We show results for point sources in red, including the effects from resolution in blue, adding line broadening in purple and marginalizing over inclinations in orange.}
\label{fig:LN_comparison}
\end{figure}

By definition, the galaxy bias in the lognormal simulations used in \textsc{Simple} is linear and the same for all galaxies. We therefore assume $b(M)=4.07$ for the comparison between the theoretical prediction and the measured $\bar{\mathcal{B}}$ from the lognormal simulations. Note that for this validation we use $\sigma_{\rm vox}^2=0.24$ computed from Eq.~\eqref{eq:sigma_vox} using also Eq.~\eqref{eq:sigma_FoG}, instead of fitting it to the measurements (see Fig.~\ref{fig:clustering} for the effect of $\sigma_{\rm vox}^2$ in the VID).  

We compare theoretical predictions and measurements from the simulations in  Fig.~\ref{fig:LN_comparison}, including $\sqrt{\mathcal{C}_{aa}}$ as a reference for the uncertainties. The model provides a good description of the mean measurements from the lognormal realizations, successfully capturing the effects of the smoothing. 

\begin{figure}[t]
\centering
\includegraphics[width=\columnwidth]{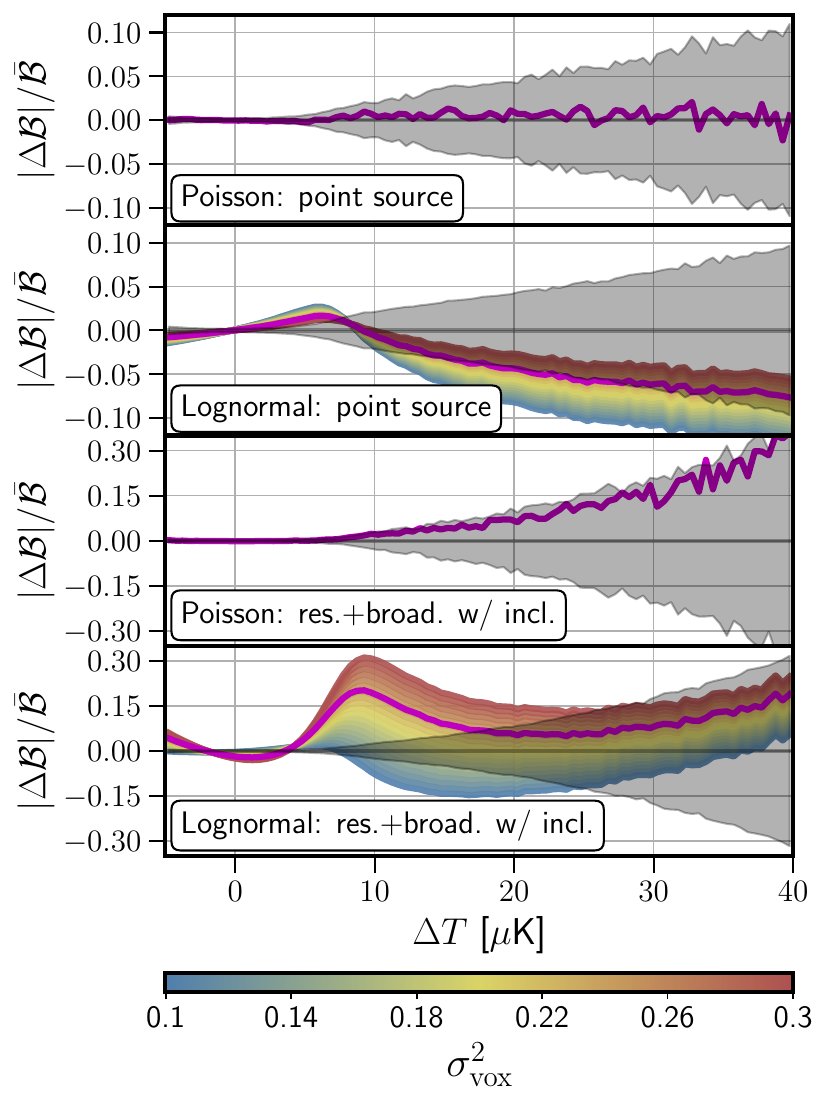}
\caption{Relative deviation between the theoretical prediction and the mean from Poisson realizations and lognormal simulations, considering point sources and extended profiles as indicated by the titles of the panels. For the cases with clustering (lognormal), we vary $\sigma_{\rm vox}^2$, highlighting the fiducial value used throughout the rest of the study in magenta. Shaded gray areas show the relative uncertainties related with the diagonal of the covariance matrix. Note the change of scale in the $y$-axis between the two top and the two bottom panels.}
\label{fig:clustering_bias}
\end{figure}

There is a small absolute deviation between the prediction and the simulations at intermediate scales, present in all four cases. We show the relative difference between the theoretical prediction and the measurement from simulations in Fig.~\ref{fig:clustering_bias}. Regarding the clustering, we consider the lognormal simulations introduced above, but also Poisson-sampled realizations without any clustering of emitters; regarding the intensity profiles, we consider point sources and extended profiles with effects from resolution and line broadening marginalizing over the inclination. The mean and covariance for the Poisson realizations are obtained from 100 independent realizations. Here we can appreciate that although our model qualitatively captures the effects of extended profiles, it does not provide an accurate prediction of the VID. The inaccuracy is most severe at intermediate brightness temperatures, where the relative deviation is maximum and the covariance is smaller.  

We associate this deviation to two potential sources of error. First, we can see that even for point sources clustering is not properly captured. This is because the modeling of clustering is limited. Truncating the moment-generating function at second order removes any impact of non Gaussianity of the halo distribution in the PDF.\footnote{Lognormal simulations exhibit a higher degree of non Gaussianity in the galaxy distribution than N-body simulations (see e.g.,~\cite{Blot:2018oxk}). Therefore, deviations from the measurements in the lognormal realizations related to the truncation of the moment expansion are expected to be larger than in a more realistic case.} This could be captured including higher terms in the expansion of the cumulants of clustering in Eq.~\eqref{eq:expansion} (and subsequently in Eq.~\eqref{eq:Fpdf_pointsource}). Note that neglecting nonlinear bias is most likely a bigger issue, but our lognormal simulations assume linear bias by definition. Second, there is a small deviation introduced by the extended profiles even in the absence of clustering. This may be due to limitations of our model or to potential numerical artifacts in the implementation of the line broadening in the simulated line-intensity maps. While the impact of artifacts related with resolution and filters in the power spectrum have been thoroughly investigated, there is significantly less studies on their effects on the histogram of the map. Furthermore, these deviations may not depend trivially on the experimental setup and ratio between $\sigma_\parallel$ and $\sigma_v$. We leave a dedicated study with a deeper study of the numerical artifacts for future work. 

Finally, we study the impact of the extended profiles in the covariance matrix of the VID. An analytic estimation of the covariance can be obtained from the 2-point PDF, following the derivation described in Ref.~\cite{Thiele:2020rig} and adapting it to LIM. However, the covariance of the VID is usually assumed to be that of a multinomial distribution, since the VID is a histogram.  A multinomial distribution presents a covariance
\begin{equation}
    \mathcal{C}_{ab}^{\rm multinom.}= \begin{cases}
    \frac{\mathcal{B}_a}{N_{\rm vox}}(1-\mathcal{B}_a)\, &{\rm if}\, a=b,\\
    -\frac{\mathcal{B}_a\mathcal{B}_b}{N_{\rm vox}}\, &{\rm if}\, a\neq b,
    \end{cases}
\end{equation}
so that its correlation matrix, defined in general as $\mathcal{R}_{ab} \equiv \mathcal{C}_{ab}/\sqrt{\mathcal{C}_{aa}\mathcal{C}_{bb}}$, has off-diagonal terms given by $\mathcal{R}_{ab}^{\rm multinom.}=-\sqrt{\mathcal{B}_a\mathcal{B}_b/[(1-\mathcal{B}_a)(1-\mathcal{B}_b)]}$.

\begin{figure}[t]
\centering
\includegraphics[width=\columnwidth]{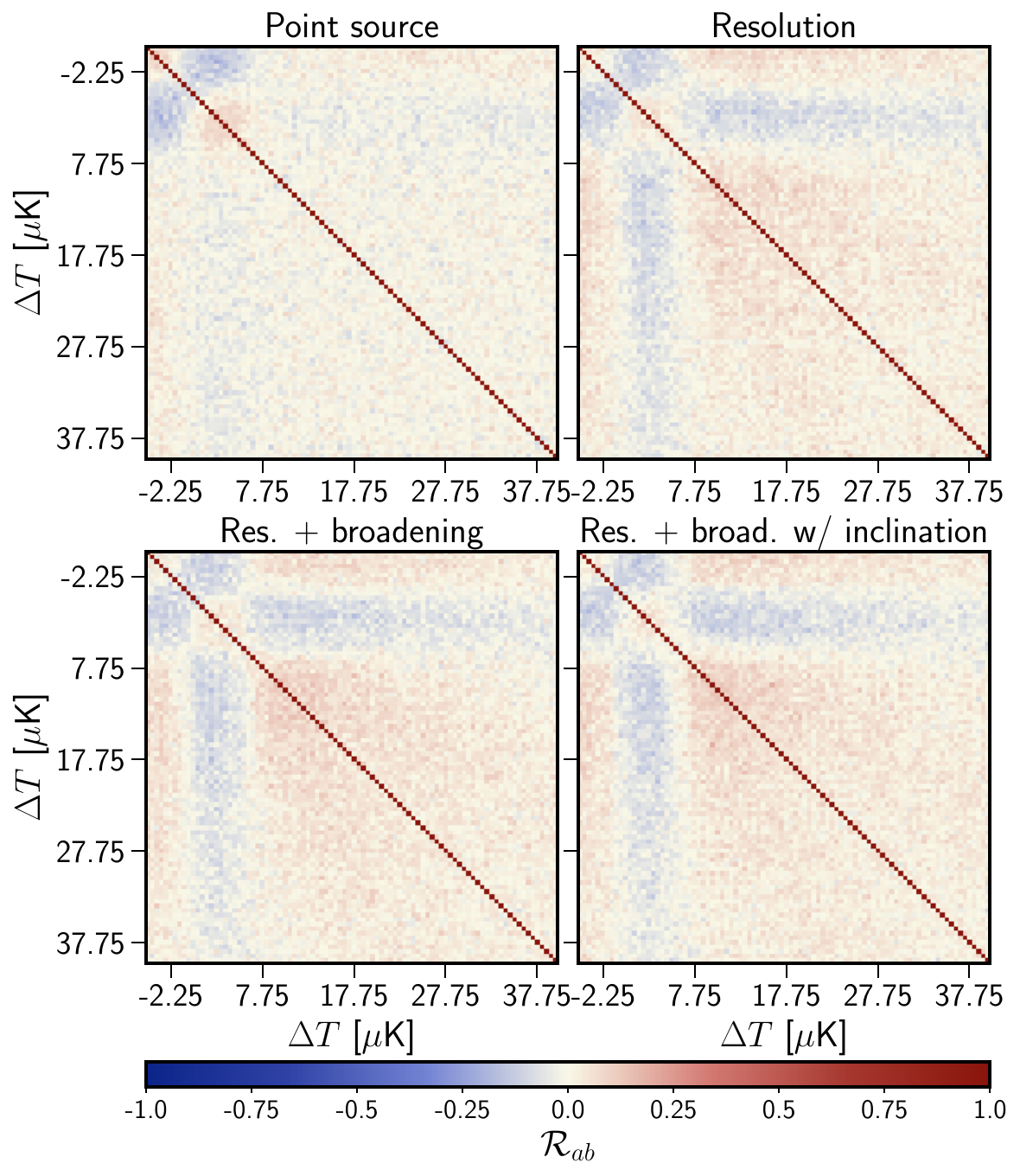}
\caption{Correlation matrices computed numerically from 1008 lognormal realizations. We consider 4 cases as indicated by the titles at the top of each panel.}
\label{fig:correlation}
\end{figure}

The diagonal of the numerical covariance computed using Eq.~\eqref{eq:covariance} agrees very well with the multinomial variance for all cases, but the off-diagonal terms do not. We show the numerical correlation matrices  computed from the lognormal realizations for the four cases in Fig.~\ref{fig:correlation}. We do not find the strong negative correlations expected for the multinomial distribution. On the contrary, we find negligible correlation for point sources except for $\Delta T\lesssim 5\,\mu{\rm K}$, which is dominated by the instrumental noise. Once emitters are allowed to contribute to more than one voxel, the off-diagonal correlation grows to $\lesssim 0.1$; it is slightly higher once line broadening is considered. The deviation from the multinomial distribution off-diagonal covariance is due to physical effects and sources contributing to several voxels inducing correlations beyond the multinomial distribution. Note that the correlation would be larger for smaller voxels, as demonstrated in Ref.~\cite{Vernstrom:2013vva}. 

The covariance measured from the simulations does not include the supersample contributions. These contributions, derived for a previous formalism of the VID in Ref.~\cite{Sato-Polito:2022fkd}, have been found to be generally smaller than statistical covariance from Poisson sampling of a histogram. Similarly, the off-diagonal physical covariance is also expected to be larger than the supersample covariance. 

\section{Parameter-inference bias}
\label{sec:bias}
We have shown that the VID accounting for extended emission is very different that the standard computation for point sources. Therefore, an analysis assuming point sources is prone to obtain highly biased results. 

We consider an analysis of the VID with the same experimental setup and model as in previous sections. We focus on the $L(M)$ relation and take $C$, $A$, $B$ and $M_\star$ as free parameters (see Eq.~\eqref{eq:LofM}). We also include $\sigma_{\rm vox}^2$ as nuisance parameter. Assuming Gaussian errors, the logarithm of the likelihood for the VID at position $\pmb{p}$ in the parameter space is
\begin{equation}
    -2 \log \mathcal{L}(\pmb{p}) = \sum_{ab}\Delta\mathcal{B}_a(\pmb{p}) \left(\mathcal{C}^{-1}\right)_{ab}\Delta\mathcal{B}_b(\pmb{p})\,,
\end{equation}
where $\Delta\mathcal{B}_a$ is the difference between the measured VID and the prediction. We use the covariance matrix numerically estimated in Sec.~\ref{sec:sims}. However, the inverse of a covariance matrix estimated from a finite number of realizations is not unbiased; we apply the correction factor described in Ref.~\cite{Hartlap:2006kj} to the inverse of the covariance matrix. In our case, the correction amounts to a $\sim 10\%$ factor.

Let us assume, for this exercise, that the data is perfectly described by our modeling of the VID including the extended signal caused by experimental resolution and line broadening for random galaxy inclinations for a set of \textit{true} parameter values $\pmb{p}_{\rm fid}$. An analysis for this model would then return best-fit parameter values $\pmb{p}_{\rm bf}^{\rm sm} = \pmb{p}_{\rm fid}$ and $\Delta\mathcal{B}_a(\pmb{p}_{\rm fid})=0$. 

For an incorrect modeling ---assuming point sources in our case---, the best-fit $\pmb{p}_{\rm bf}^{\rm ps}$ inferred will be biased. This bias can be estimated using a Fisher-matrix analysis, through a linear approximation of the likelihood. This is not a good approximation for our case, since we expect large deviations in $\mathcal{B}$ for the incorrect model (see Fig.~\ref{fig:VID}). However, this exercise provides a qualitative estimate of the impact in parameter inference. 

We denote the predictions for the two models with $\mathcal{B}^{\rm sm}$ and $\mathcal{B}^{\rm ps}$, respectively, standing for `smoothed' and `point sources'. The estimated systematic bias introduced due to the poor modeling is~\cite{Bernal:2020pwq}
\begin{equation}
\begin{split}
    & \Delta\pmb{p}_{\rm syst}  \equiv \pmb{p}_{\rm bf}^{\rm ps}-\pmb{p}_{\rm fid} = \\
    & = F_{\rm ps}^{-1}\sum_{ab}\pmb{\nabla}_{p}\mathcal{B}_a^{\rm ps}\vert_{\pmb{p}_{\rm fid}} \left(\mathcal{C}_{\rm ps}^{-1}\right)_{ab}\left[\mathcal{B}_b^{\rm sm}(\pmb{p}_{\rm fid})-\mathcal{B}_b^{\rm ps}(\pmb{p}_{\rm fid})\right]\,,
\end{split}
\label{eq:syst_bias}
\end{equation}
where 
\begin{equation}
    F = \sum_{ab}\pmb{\nabla}_{p}\mathcal{B}_a\vert_{\pmb{p}_{\rm fid}}\left(\mathcal{C}^{-1}\right)_{ab}\left(\pmb{\nabla}_{p}\mathcal{B}_b\vert_{\pmb{p}_{\rm fid}}\right)
\end{equation}
is the Fisher matrix computed for the case of interest, and $\pmb{\nabla}_{p}\mathcal{B}_a^{\rm ps}\vert_{\pmb{p}_{\rm fid}}$ is the gradient in parameter space of the VID evaluated at $\pmb{p}_{\rm fid}$.

\begin{figure}[t]
\centering
\includegraphics[width=\linewidth]{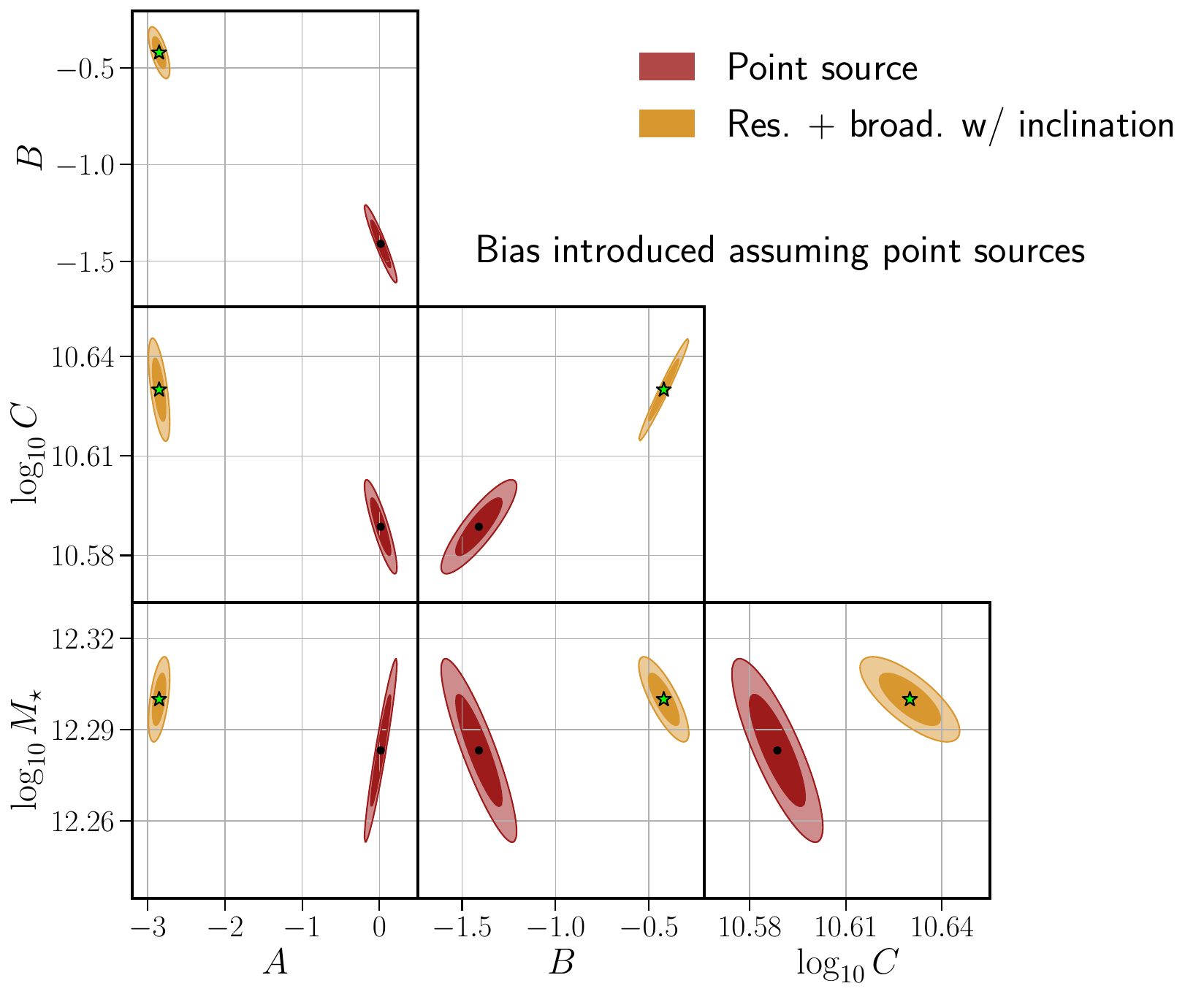}
\caption{68\% and 95\% confidence level forecasted marginalized constraints on the parameters controlling the $L(M)$ relation for the experimental setup described in Sec.~\ref{sec:exp_astro} and~\ref{sec:sims}. We compare the results for point sources (red) and modeling the limited resolution of the experiment and line broadening marginalizing over the emitter inclination (orange), assuming that the latter is the correct description of the data. \textit{True} parameter values are marked by a star, while the best fit assuming point sources is marked by a black dot.}
\label{fig:fisher_bias}
\end{figure}

We show marginalized forecasted uncertainties for the two models including the estimation for the systematic bias in Fig.~\ref{fig:fisher_bias}. All parameters present big systematic biases, except for $\log_{10}M_\star$, for which the marginalized 1-dimensional bias is $\sim 1.8\sigma$. In addition, note that the constraining power and the degeneracies obtained if point sources are assumed are also different. Incomplete modeling may affect the conclusions obtained from a Fisher forecasts, which motivates the use of more accurate models even for qualitative forecast sensitivity~\cite{Bellomo:2020pnw}.

\section{Conclusions}
\label{sec:conclusions}
Line-intensity mapping proposes an alternative tracer for the large-scale structure of the Universe, providing unprecedented sensitivity to high redshifts and faint emitters. Contrary to other tracers, LIM is also sensitive to astrophysical processes driving galaxy formation and evolution. This additional dependence adds another layer of complexity to the line-intensity fluctuation maps, which become very non Gaussian. Therefore, LIM power spectra can only capture a fraction of the information encoded in line-intensity maps. Furthermore, the power spectrum is affected by degeneracies between cosmology and astrophysics. These are the main motivations to explore one-point statistics, since they probe the whole distribution of intensity fluctuations and are more sensitive to features in the line-luminosity function.

So far, the modeling of LIM one-point statistics has relied on an unphysical assumption: that each emitter contributes just to the voxel that contains it. This assumption implies that experiments have perfect resolution and that all emitters are point sources with negligible line width.

In this work we have dropped this assumption and implemented for the first time the contribution of an emitter to several voxels in the LIM PDF theoretical prediction. We build upon previous work which use the halo model~\cite{Thiele:2018jdl,Thiele:2020rig,Breysse:2022alx}, noting the additional complication that LIM implies with respect to projected angular fields. We circumvent this complication invoking the Ergodic hypothesis and integrating over the spatial distribution of the intensity profile when computing the intensity PDF of a single emitter. 

We compare four different cases: point sources, adding the effects of limited resolution, adding on top it line broadening assuming that all galaxies are perfectly edge-on, and marginalizing over their inclination. Extended profiles significantly alter the LIM PDF prediction, with changes of more than one order of magnitudes. The resulting PDF has a broader, lower peak at low intensities and a lower, shorter tail towards high intensities. This follows our intuition: if an emitter contributes to several voxels, there will be fewer empty or very faint voxels and also less very bright voxels, respectively. We find that, for the example considered, the dominant effect is related to the limited resolution, while including the effects of line broadening adds a small modification. Since line broadening only acts in the direction along the line of sight and is more relevant for massive, rare emitters, it has a smaller effect on the VID than the the experimental resolution, which acts in the three directions and affects all emitters equally. The qualitative behaviour agrees with our expectations, and although the LIM PDF depends a lot on the specific astrophysical model and experimental setup, we expect these results to be generally applicable. 

We validate, for the first time, the theoretical prediction of the VID against simulations. We compare our implementation with lognormal simulations and find good qualitative agreement. Nonetheless, we still find deviations that are related with the clustering of emitters, as well as potential numerical artifacts from the implementation of line broadening in the simulations or limitations in our model to include extended profiles in the theoretical prediction. We leave further development of the modeling of clustering for the LIM PDF and a deeper study of the implementation of extended profiles in simulations and its impact on the measured VID for future work. Alternative frameworks like the theory of large deviations, which is being developed for the PDF of cosmic magnification (see e.g. Ref.~\cite{Barthelemy:2023mer} and references therein), provide a better prediction for the matter clustering PDF. Therefore, a combination of both approaches, for instance combining the LIM PDF for unclustered sources with a PDF for the clustering of such sources, might lead to a more accurate prediction.

We also study the effects of extended profiles in the VID covariance. First, we see that while the diagonal of the covariance follows closely the variance for a multinomial distribution, as expected from a histogram, the off-diagonal terms significantly deviate. We find negligible correlation for point sources, instead of the strong negative correlations expected for a multinomial distribution. Second, we find that extended profiles increase the correlation between intensity bins in the VID, reaching correlations of around 10\%. Finally, we demonstrate that very large systematic biases in parameter inference can be expected if point-like intensity profiles are assumed.

One-point statistics, especially through its complementarity with the power spectrum, holds promise to break parameter degeneracies and significantly increase the constraining power of LIM surveys. In the case observational contaminants preclude the use of the VID, there are alternatives, combining different LIM observations or line-intensity maps and galaxy catalogs, which are more robust to observational systematics. However, since these alternative summary statistics still depend on the intensity PDF, they must account for the profiles of the observed signal from each emitter. 

This work is but one of the first steps towards an accurate and efficient modeling for the LIM PDF. Further research is required to achieve a successful application to observations. Nonetheless, the promise of LIM one-point summary statistics motivates this effort, since only through their combination with LIM power spectra and other observations LIM surveys will unlock their full potential. 

\begin{acknowledgments}
The author thanks Nickolas Kokron and Gabriela Sato-Polito for key discussions during the development of this work, and especially Patrick Breysse for important discussions and also sharing the implementation of the LIM PDF from his previous work. The author acknowledges funding from the Ramón y Cajal Grant RYC2021-033191-I, financed by MCIN/AEI/10.13039/501100011033 and by the European Union “NextGenerationEU”/PRTR, and the computer resources provided by the Spanish Supercomputing Network (RES) node at Universidad de Cantabria and the Institute of Physics of Cantabria (IFCA-CSIC). 
\end{acknowledgments}

\bibliography{Refs.bib}

\providecommand{\href}[2]{#2}\begingroup\raggedright\begin{thebibliography}{10}

\bibitem{Kovetz:2017agg}
E.~D. Kovetz {\em et~al.}, ``{Line-Intensity Mapping: 2017 Status Report},''
  \href{http://arxiv.org/abs/1709.09066}{{\ttfamily arXiv:1709.09066
  [astro-ph.CO]}}.

\bibitem{2012RPPh...75h6901P}
J.~R. {Pritchard} and A.~{Loeb}, ``{21 cm cosmology in the 21st century},''
  \href{http://dx.doi.org/10.1088/0034-4885/75/8/086901}{{\em Reports on
  Progress in Physics} {\bfseries 75} no.~8, (Aug., 2012) 086901},
  \href{http://arxiv.org/abs/1109.6012}{{\ttfamily arXiv:1109.6012
  [astro-ph.CO]}}.

\bibitem{Liu:2019awk}
A.~Liu and J.~R. Shaw, ``{Data Analysis for Precision 21 cm Cosmology},''
  \href{http://dx.doi.org/10.1088/1538-3873/ab5bfd}{{\em Publ. Astron. Soc.
  Pac.} {\bfseries 132} no.~1012, (2020) 062001},
  \href{http://arxiv.org/abs/1907.08211}{{\ttfamily arXiv:1907.08211
  [astro-ph.IM]}}.

\bibitem{Bernal:2022jap}
J.~L. Bernal and E.~D. Kovetz, ``{Line-intensity mapping: theory review with a
  focus on star-formation lines},''
  \href{http://dx.doi.org/10.1007/s00159-022-00143-0}{{\em Astron. Astrophys.
  Rev.} {\bfseries 30} no.~1, (2022) 5},
  \href{http://arxiv.org/abs/2206.15377}{{\ttfamily arXiv:2206.15377
  [astro-ph.CO]}}.

\bibitem{Cheng:2018hox}
Y.-T. Cheng, R.~de~Putter, T.-C. Chang, and O.~Dore, ``{Optimally Mapping
  Large-Scale Structures with Luminous Sources},''
  \href{http://dx.doi.org/10.3847/1538-4357/ab1b2b}{{\em Astrophys. J.}
  {\bfseries 877} no.~2, (2019) 86},
  \href{http://arxiv.org/abs/1809.06384}{{\ttfamily arXiv:1809.06384
  [astro-ph.CO]}}.

\bibitem{Schaan:2021hhy}
E.~Schaan and M.~White, ``{Astrophysics \& Cosmology from Line Intensity
  Mapping vs Galaxy Surveys},''
  \href{http://dx.doi.org/10.1088/1475-7516/2021/05/067}{{\em JCAP} {\bfseries
  05} (2021) 067}, \href{http://arxiv.org/abs/2103.01971}{{\ttfamily
  arXiv:2103.01971 [astro-ph.CO]}}.

\bibitem{vanHaarlem:2013dsa}
M.~P. van Haarlem {\em et~al.}, ``{LOFAR: The LOw-Frequency ARray},''
  \href{http://dx.doi.org/10.1051/0004-6361/201220873}{{\em Astron. Astrophys.}
  {\bfseries 556} (2013) A2}, \href{http://arxiv.org/abs/1305.3550}{{\ttfamily
  arXiv:1305.3550 [astro-ph.IM]}}.

\bibitem{Bandura:2014gwa}
K.~Bandura {\em et~al.}, ``{Canadian Hydrogen Intensity Mapping Experiment
  (CHIME) Pathfinder},'' \href{http://dx.doi.org/10.1117/12.2054950}{{\em Proc.
  SPIE Int. Soc. Opt. Eng.} {\bfseries 9145} (2014) 22},
  \href{http://arxiv.org/abs/1406.2288}{{\ttfamily arXiv:1406.2288
  [astro-ph.IM]}}.

\bibitem{DeBoer:2016tnn}
D.~R. DeBoer {\em et~al.}, ``{Hydrogen Epoch of Reionization Array (HERA)},''
  \href{http://dx.doi.org/10.1088/1538-3873/129/974/045001}{{\em Publ. Astron.
  Soc. Pac.} {\bfseries 129} no.~974, (2017) 045001},
  \href{http://arxiv.org/abs/1606.07473}{{\ttfamily arXiv:1606.07473
  [astro-ph.IM]}}.

\bibitem{MeerKLASS:2017vgf}
{\bfseries MeerKLASS} Collaboration, M.~G. Santos {\em et~al.}, ``{MeerKLASS:
  MeerKAT Large Area Synoptic Survey},'' in {\em {MeerKAT Science}: {On the
  Pathway to the SKA}}.
\newblock 9, 2017.
\newblock \href{http://arxiv.org/abs/1709.06099}{{\ttfamily arXiv:1709.06099
  [astro-ph.CO]}}.

\bibitem{Cleary:2021dsp}
K.~A. Cleary {\em et~al.}, ``{COMAP Early Science: I. Overview},''
  \href{http://arxiv.org/abs/2111.05927}{{\ttfamily arXiv:2111.05927
  [astro-ph.CO]}}.

\bibitem{CONCERTO:2020ahk}
{\bfseries CONCERTO} Collaboration, P.~Ade {\em et~al.}, ``{A wide
  field-of-view low-resolution spectrometer at APEX: Instrument design and
  scientific forecast},''
  \href{http://dx.doi.org/10.1051/0004-6361/202038456}{{\em Astron. Astrophys.}
  {\bfseries 642} (2020) A60},
  \href{http://arxiv.org/abs/2007.14246}{{\ttfamily arXiv:2007.14246
  [astro-ph.IM]}}.

\bibitem{Gebhardt:2021vfo}
K.~Gebhardt {\em et~al.}, ``{The Hobby\textendash{}Eberly Telescope Dark Energy
  Experiment (HETDEX) Survey Design, Reductions, and Detections*},''
  \href{http://dx.doi.org/10.3847/1538-4357/ac2e03}{{\em Astrophys. J.}
  {\bfseries 923} no.~2, (2021) 217},
  \href{http://arxiv.org/abs/2110.04298}{{\ttfamily arXiv:2110.04298
  [astro-ph.IM]}}.

\bibitem{CCAT-Prime:2021lly}
{\bfseries CCAT-Prime} Collaboration, M.~Aravena {\em et~al.}, ``{CCAT-prime
  Collaboration: Science Goals and Forecasts with Prime-Cam on the Fred Young
  Submillimeter Telescope},''
  \href{http://dx.doi.org/10.3847/1538-4365/ac9838}{{\em Astrophys. J. Suppl.}
  {\bfseries 264} no.~1, (2023) 7},
  \href{http://arxiv.org/abs/2107.10364}{{\ttfamily arXiv:2107.10364
  [astro-ph.CO]}}.

\bibitem{Sun:2020mco}
G.~Sun {\em et~al.}, ``{Probing Cosmic Reionization and Molecular Gas Growth
  with TIME},'' \href{http://dx.doi.org/10.3847/1538-4357/abfe62}{{\em
  Astrophys. J.} {\bfseries 915} no.~1, (2021) 33},
  \href{http://arxiv.org/abs/2012.09160}{{\ttfamily arXiv:2012.09160
  [astro-ph.GA]}}.

\bibitem{Switzer:2021jeg}
E.~R. Switzer {\em et~al.}, ``{Experiment for cryogenic large-aperture
  intensity mapping: instrument design},''
  \href{http://dx.doi.org/10.1117/1.JATIS.7.4.044004}{{\em J. Astron. Telesc.
  Instrum. Syst.} {\bfseries 7} no.~4, (2021) 044004}.

\bibitem{2020arXiv200914340V}
J.~{Vieira}, J.~{Aguirre}, C.~M. {Bradford}, J.~{Filippini}, C.~{Groppi},
  D.~{Marrone}, {\em et~al.}, ``{The Terahertz Intensity Mapper (TIM): a
  Next-Generation Experiment for Galaxy Evolution Studies},''
  \href{http://dx.doi.org/10.48550/arXiv.2009.14340}{{\em arXiv e-prints}
  (Sept., 2020) arXiv:2009.14340},
  \href{http://arxiv.org/abs/2009.14340}{{\ttfamily arXiv:2009.14340
  [astro-ph.IM]}}.

\bibitem{Dore:2014cca}
O.~Dor\'e {\em et~al.}, ``{Cosmology with the SPHEREX All-Sky Spectral
  Survey},'' \href{http://arxiv.org/abs/1412.4872}{{\ttfamily arXiv:1412.4872
  [astro-ph.CO]}}.

\bibitem{Koopmans:2015sua}
L.~V.~E. Koopmans {\em et~al.}, ``{The Cosmic Dawn and Epoch of Reionization
  with the Square Kilometre Array},''
  \href{http://dx.doi.org/10.22323/1.215.0001}{{\em PoS} {\bfseries AASKA14}
  (2015) 001}, \href{http://arxiv.org/abs/1505.07568}{{\ttfamily
  arXiv:1505.07568 [astro-ph.CO]}}.

\bibitem{Newburgh:2016mwi}
L.~B. Newburgh {\em et~al.}, ``{HIRAX: A Probe of Dark Energy and Radio
  Transients},'' \href{http://dx.doi.org/10.1117/12.2234286}{{\em Proc. SPIE
  Int. Soc. Opt. Eng.} {\bfseries 9906} (2016) 99065X},
  \href{http://arxiv.org/abs/1607.02059}{{\ttfamily arXiv:1607.02059
  [astro-ph.IM]}}.

\bibitem{Chang:2007xk}
T.-C. Chang, U.-L. Pen, J.~B. Peterson, and P.~McDonald, ``{Baryon Acoustic
  Oscillation Intensity Mapping as a Test of Dark Energy},''
  \href{http://dx.doi.org/10.1103/PhysRevLett.100.091303}{{\em Phys. Rev.
  Lett.} {\bfseries 100} (2008) 091303},
  \href{http://arxiv.org/abs/0709.3672}{{\ttfamily arXiv:0709.3672
  [astro-ph]}}.

\bibitem{Keating:2016pka}
G.~K. Keating, D.~P. Marrone, G.~C. Bower, E.~Leitch, J.~E. Carlstrom, and
  D.~R. DeBoer, ``{COPSS II: The molecular gas content of ten million cubic
  megaparsecs at redshift z \ensuremath{\sim} 3},''
  \href{http://dx.doi.org/10.3847/0004-637X/830/1/34}{{\em Astrophys. J.}
  {\bfseries 830} no.~1, (2016) 34},
  \href{http://arxiv.org/abs/1605.03971}{{\ttfamily arXiv:1605.03971
  [astro-ph.GA]}}.

\bibitem{Yang:2019eoj}
S.~Yang, A.~R. Pullen, and E.~R. Switzer, ``{Evidence for C II diffuse line
  emission at redshift $z\sim2.6$},''
  \href{http://dx.doi.org/10.1093/mnrasl/slz126}{{\em Mon. Not. Roy. Astron.
  Soc.} {\bfseries 489} no.~1, (2019) L53--L57},
  \href{http://arxiv.org/abs/1904.01180}{{\ttfamily arXiv:1904.01180
  [astro-ph.CO]}}.

\bibitem{Keating:2020wlx}
G.~K. Keating, D.~P. Marrone, G.~C. Bower, and R.~P. Keenan, ``{An Intensity
  Mapping Detection of Aggregate CO Line Emission at 3 mm},''
  \href{http://dx.doi.org/10.3847/1538-4357/abb08e}{{\em Astrophys. J.}
  {\bfseries 901} no.~2, (2020) 141},
  \href{http://arxiv.org/abs/2008.08087}{{\ttfamily arXiv:2008.08087
  [astro-ph.GA]}}.

\bibitem{Wolz:2021ofa}
L.~Wolz {\em et~al.}, ``{H\,i constraints from the cross-correlation of eBOSS
  galaxies and Green Bank Telescope intensity maps},''
  \href{http://dx.doi.org/10.1093/mnras/stab3621}{{\em Mon. Not. Roy. Astron.
  Soc.} {\bfseries 510} no.~3, (2022) 3495--3511},
  \href{http://arxiv.org/abs/2102.04946}{{\ttfamily arXiv:2102.04946
  [astro-ph.CO]}}.

\bibitem{Niemeyer:2022vrt}
M.~L. Niemeyer {\em et~al.}, ``{Surface Brightness Profile of
  Lyman-\ensuremath{\alpha} Halos out to 320 kpc in HETDEX},''
  \href{http://dx.doi.org/10.3847/1538-4357/ac5cb8}{{\em Astrophys. J.}
  {\bfseries 929} no.~1, (2022) 90},
  \href{http://arxiv.org/abs/2203.04826}{{\ttfamily arXiv:2203.04826
  [astro-ph.GA]}}.

\bibitem{Cunnington:2022uzo}
S.~Cunnington {\em et~al.}, ``{H\,i intensity mapping with MeerKAT: power
  spectrum detection in cross-correlation with WiggleZ galaxies},''
  \href{http://dx.doi.org/10.1093/mnras/stac3060}{{\em Mon. Not. Roy. Astron.
  Soc.} {\bfseries 518} no.~4, (2022) 6262--6272},
  \href{http://arxiv.org/abs/2206.01579}{{\ttfamily arXiv:2206.01579
  [astro-ph.CO]}}.

\bibitem{Niemeyer:2022arn}
M.~L. Niemeyer {\em et~al.}, ``{Ly\ensuremath{\alpha} Halos around [O
  iii]-selected Galaxies in HETDEX},''
  \href{http://dx.doi.org/10.3847/2041-8213/ac82e5}{{\em Astrophys. J. Lett.}
  {\bfseries 934} no.~2, (2022) L26},
  \href{http://arxiv.org/abs/2207.11098}{{\ttfamily arXiv:2207.11098
  [astro-ph.GA]}}.

\bibitem{Paul:2023yrr}
S.~Paul, M.~G. Santos, Z.~Chen, and L.~Wolz, ``{A first detection of neutral
  hydrogen intensity mapping on Mpc scales at $z\approx 0.32$ and $z\approx
  0.44$},'' \href{http://arxiv.org/abs/2301.11943}{{\ttfamily arXiv:2301.11943
  [astro-ph.CO]}}.

\bibitem{MoradinezhadDizgah:2018lac}
A.~Moradinezhad~Dizgah and G.~K. Keating, ``{Line intensity mapping with [CII]
  and CO(1-0) as probes of primordial non-Gaussianity},''
  \href{http://dx.doi.org/10.3847/1538-4357/aafd36}{{\em Astrophys. J.}
  {\bfseries 872} no.~2, (2019) 126},
  \href{http://arxiv.org/abs/1810.02850}{{\ttfamily arXiv:1810.02850
  [astro-ph.CO]}}.

\bibitem{Bernal:2019gfq}
J.~L. Bernal, P.~C. Breysse, and E.~D. Kovetz, ``{Cosmic Expansion History from
  Line-Intensity Mapping},''
  \href{http://dx.doi.org/10.1103/PhysRevLett.123.251301}{{\em Phys. Rev.
  Lett.} {\bfseries 123} no.~25, (2019) 251301},
  \href{http://arxiv.org/abs/1907.10065}{{\ttfamily arXiv:1907.10065
  [astro-ph.CO]}}.

\bibitem{Sato-Polito:2020cil}
G.~Sato-Polito, J.~L. Bernal, K.~K. Boddy, and M.~Kamionkowski, ``{Kinetic
  Sunyaev-Zel\textquoteright{}dovich tomography with line-intensity mapping},''
  \href{http://dx.doi.org/10.1103/PhysRevD.103.083519}{{\em Phys. Rev. D}
  {\bfseries 103} no.~8, (2021) 083519},
  \href{http://arxiv.org/abs/2011.08193}{{\ttfamily arXiv:2011.08193
  [astro-ph.CO]}}.

\bibitem{MoradinezhadDizgah:2021upg}
A.~Moradinezhad~Dizgah, G.~K. Keating, K.~S. Karkare, A.~Crites, and S.~R.
  Choudhury, ``{Neutrino Properties with Ground-based Millimeter-wavelength
  Line Intensity Mapping},''
  \href{http://dx.doi.org/10.3847/1538-4357/ac3edd}{{\em Astrophys. J.}
  {\bfseries 926} no.~2, (2022) 137},
  \href{http://arxiv.org/abs/2110.00014}{{\ttfamily arXiv:2110.00014
  [astro-ph.CO]}}.

\bibitem{Castorina:2019zho}
E.~Castorina and M.~White, ``{Measuring the growth of structure with intensity
  mapping surveys},''
  \href{http://dx.doi.org/10.1088/1475-7516/2019/06/025}{{\em JCAP} {\bfseries
  06} (2019) 025}, \href{http://arxiv.org/abs/1902.07147}{{\ttfamily
  arXiv:1902.07147 [astro-ph.CO]}}.

\bibitem{MoradinezhadDizgah:2021dei}
A.~Moradinezhad~Dizgah, F.~Nikakhtar, G.~K. Keating, and E.~Castorina,
  ``{Precision tests of CO and CII power spectra models against simulated
  intensity maps},''
  \href{http://dx.doi.org/10.1088/1475-7516/2022/02/026}{{\em JCAP} {\bfseries
  02} no.~02, (2022) 026}, \href{http://arxiv.org/abs/2111.03717}{{\ttfamily
  arXiv:2111.03717 [astro-ph.CO]}}.

\bibitem{Bernal:2019jdo}
J.~L. Bernal, P.~C. Breysse, H.~Gil-Mar\'\i{}n, and E.~D. Kovetz,
  ``{User\textquoteright{}s guide to extracting cosmological information from
  line-intensity maps},''
  \href{http://dx.doi.org/10.1103/PhysRevD.100.123522}{{\em Phys. Rev. D}
  {\bfseries 100} no.~12, (2019) 123522},
  \href{http://arxiv.org/abs/1907.10067}{{\ttfamily arXiv:1907.10067
  [astro-ph.CO]}}.

\bibitem{Camera:2019iwy}
S.~Camera and H.~Padmanabhan, ``{Beyond \ensuremath{\Lambda}CDM with H i
  intensity mapping: robustness of cosmological constraints in the presence of
  astrophysics},'' \href{http://dx.doi.org/10.1093/mnras/staa1663}{{\em Mon.
  Not. Roy. Astron. Soc.} {\bfseries 496} no.~4, (2020) 4115--4126},
  \href{http://arxiv.org/abs/1910.00022}{{\ttfamily arXiv:1910.00022
  [astro-ph.CO]}}.

\bibitem{Breysse:2015saa}
P.~C. Breysse, E.~D. Kovetz, and M.~Kamionkowski, ``{The high redshift
  star-formation history from carbon-monoxide intensity maps},''
  \href{http://dx.doi.org/10.1093/mnrasl/slw005}{{\em Mon. Not. Roy. Astron.
  Soc.} {\bfseries 457} no.~1, (2016) L127--L131},
  \href{http://arxiv.org/abs/1507.06304}{{\ttfamily arXiv:1507.06304
  [astro-ph.CO]}}.

\bibitem{Breysse:2016szq}
P.~C. Breysse, E.~D. Kovetz, P.~S. Behroozi, L.~Dai, and M.~Kamionkowski,
  ``{Insights from probability distribution functions of intensity maps},''
  \href{http://dx.doi.org/10.1093/mnras/stx203}{{\em Mon. Not. Roy. Astron.
  Soc.} {\bfseries 467} no.~3, (2017) 2996--3010},
  \href{http://arxiv.org/abs/1609.01728}{{\ttfamily arXiv:1609.01728
  [astro-ph.CO]}}.

\bibitem{COMAP:2018kem}
{\bfseries COMAP} Collaboration, H.~T. Ihle {\em et~al.}, ``{Joint power
  spectrum and voxel intensity distribution forecast on the CO luminosity
  function with COMAP},''
  \href{http://dx.doi.org/10.3847/1538-4357/aaf4bc}{{\em Astrophys. J.}
  {\bfseries 871} no.~1, (2019) 75},
  \href{http://arxiv.org/abs/1808.07487}{{\ttfamily arXiv:1808.07487
  [astro-ph.CO]}}.

\bibitem{Sato-Polito:2022fkd}
G.~Sato-Polito and J.~L. Bernal, ``{Analytical covariance between the voxel
  intensity distribution and the power spectrum of line-intensity maps},''
  \href{http://dx.doi.org/10.1103/PhysRevD.106.103534}{{\em Phys. Rev. D}
  {\bfseries 106} no.~10, (2022) 103534},
  \href{http://arxiv.org/abs/2202.02330}{{\ttfamily arXiv:2202.02330
  [astro-ph.CO]}}.

\bibitem{Bernal:2020lkd}
J.~L. Bernal, A.~Caputo, and M.~Kamionkowski, ``{Strategies to Detect
  Dark-Matter Decays with Line-Intensity Mapping},''
  \href{http://dx.doi.org/10.1103/PhysRevD.103.063523}{{\em Phys. Rev. D}
  {\bfseries 103} no.~6, (2021) 063523},
  \href{http://arxiv.org/abs/2012.00771}{{\ttfamily arXiv:2012.00771
  [astro-ph.CO]}}. [Erratum: Phys.Rev.D 105, 089901 (2022)].

\bibitem{Bernal:2021ylz}
J.~L. Bernal, A.~Caputo, F.~Villaescusa-Navarro, and M.~Kamionkowski,
  ``{Searching for the Radiative Decay of the Cosmic Neutrino Background with
  Line-Intensity Mapping},''
  \href{http://dx.doi.org/10.1103/PhysRevLett.127.131102}{{\em Phys. Rev.
  Lett.} {\bfseries 127} no.~13, (2021) 131102},
  \href{http://arxiv.org/abs/2103.12099}{{\ttfamily arXiv:2103.12099
  [hep-ph]}}.

\bibitem{Bernal:2022wsu}
J.~L. Bernal, G.~Sato-Polito, and M.~Kamionkowski, ``{Cosmic Optical Background
  Excess, Dark Matter, and Line-Intensity Mapping},''
  \href{http://dx.doi.org/10.1103/PhysRevLett.129.231301}{{\em Phys. Rev.
  Lett.} {\bfseries 129} no.~23, (2022) 231301},
  \href{http://arxiv.org/abs/2203.11236}{{\ttfamily arXiv:2203.11236
  [astro-ph.CO]}}.

\bibitem{Libanore:2022ntl}
S.~Libanore, C.~Unal, D.~Sarkar, and E.~D. Kovetz, ``{Unveiling cosmological
  information on small scales with line intensity mapping},''
  \href{http://dx.doi.org/10.1103/PhysRevD.106.123512}{{\em Phys. Rev. D}
  {\bfseries 106} no.~12, (2022) 123512},
  \href{http://arxiv.org/abs/2208.01658}{{\ttfamily arXiv:2208.01658
  [astro-ph.CO]}}.

\bibitem{Adi:2023qdf}
T.~Adi, S.~Libanore, H.~A.~G. Cruz, and E.~D. Kovetz, ``{Constraining
  Primordial Magnetic Fields with Line-Intensity Mapping},''
  \href{http://arxiv.org/abs/2305.06440}{{\ttfamily arXiv:2305.06440
  [astro-ph.CO]}}.

\bibitem{1957PCPS...53..764S}
P.~A.~G. {Scheuer}, ``{A statistical method for analysing observations of faint
  radio stars},'' \href{http://dx.doi.org/10.1017/S0305004100032825}{{\em
  Proceedings of the Cambridge Philosophical Society} {\bfseries 53} (Jan.,
  1957) 764--773}.

\bibitem{1992ApJ...396..460B}
X.~{Barcons}, ``{Confusion Noise and Source Clustering},''
  \href{http://dx.doi.org/10.1086/171733}{{\em \apj} {\bfseries 396} (Sept.,
  1992) 460}.

\bibitem{Lee:2008fm}
S.~K. Lee, S.~Ando, and M.~Kamionkowski, ``{The Gamma-Ray-Flux Probability
  Distribution Function from Galactic Halo Substructure},''
  \href{http://dx.doi.org/10.1088/1475-7516/2009/07/007}{{\em JCAP} {\bfseries
  07} (2009) 007}, \href{http://arxiv.org/abs/0810.1284}{{\ttfamily
  arXiv:0810.1284 [astro-ph]}}.

\bibitem{2009ApJ...707.1750P}
G.~{Patanchon}, P.~A.~R. {Ade}, J.~J. {Bock}, E.~L. {Chapin}, M.~J. {Devlin},
  S.~R. {Dicker}, {\em et~al.}, ``{Submillimeter Number Counts from Statistical
  Analysis of BLAST Maps},''
  \href{http://dx.doi.org/10.1088/0004-637X/707/2/1750}{{\em \apj} {\bfseries
  707} no.~2, (Dec., 2009) 1750--1765},
  \href{http://arxiv.org/abs/0906.0981}{{\ttfamily arXiv:0906.0981
  [astro-ph.CO]}}.

\bibitem{Breysse:2022alx}
P.~C. Breysse, ``{Breaking the intensity-bias degeneracy in line intensity
  mapping},'' \href{http://arxiv.org/abs/2209.01223}{{\ttfamily
  arXiv:2209.01223 [astro-ph.CO]}}.

\bibitem{Thiele:2018jdl}
L.~Thiele, J.~C. Hill, and K.~M. Smith, ``{Accurate analytic model for the
  thermal Sunyaev-Zel\textquoteright{}dovich one-point probability distribution
  function},'' \href{http://dx.doi.org/10.1103/PhysRevD.99.103511}{{\em Phys.
  Rev. D} {\bfseries 99} no.~10, (2019) 103511},
  \href{http://arxiv.org/abs/1812.05584}{{\ttfamily arXiv:1812.05584
  [astro-ph.CO]}}.

\bibitem{Thiele:2020rig}
L.~Thiele, J.~C. Hill, and K.~M. Smith, ``{Accurate analytic model for the weak
  lensing convergence one-point probability distribution function and its
  autocovariance},'' \href{http://dx.doi.org/10.1103/PhysRevD.102.123545}{{\em
  Phys. Rev. D} {\bfseries 102} no.~12, (2020) 123545},
  \href{http://arxiv.org/abs/2009.06547}{{\ttfamily arXiv:2009.06547
  [astro-ph.CO]}}.

\bibitem{Breysse:2019cdw}
P.~C. Breysse, C.~J. Anderson, and P.~Berger, ``{Canceling out intensity
  mapping foregrounds},''
  \href{http://dx.doi.org/10.1103/PhysRevLett.123.231105}{{\em Phys. Rev.
  Lett.} {\bfseries 123} no.~23, (2019) 231105},
  \href{http://arxiv.org/abs/1907.04369}{{\ttfamily arXiv:1907.04369
  [astro-ph.CO]}}.

\bibitem{Breysse:2022fdi}
P.~C. Breysse, D.~T. Chung, and H.~T. Ihle, ``{Characteristic Functions for
  Cosmological Cross-Correlations},''
  \href{http://arxiv.org/abs/2210.14902}{{\ttfamily arXiv:2210.14902
  [astro-ph.CO]}}.

\bibitem{COMAP:2022sdg}
{\bfseries COMAP} Collaboration, D.~T. Chung, I.~Bangari, P.~C. Breysse, H.~T.
  Ihle, J.~R. Bond, D.~A. Dunne, {\em et~al.}, ``{The deconvolved distribution
  estimator: enhancing reionization-era CO line-intensity mapping analyses with
  a cross-correlation analogue for one-point statistics},''
  \href{http://dx.doi.org/10.1093/mnras/stad359}{{\em Mon. Not. Roy. Astron.
  Soc.} {\bfseries 520} no.~4, (2023) 5305--5316},
  \href{http://arxiv.org/abs/2210.14890}{{\ttfamily arXiv:2210.14890
  [astro-ph.CO]}}.

\bibitem{Planck:2018vyg}
{\bfseries Planck} Collaboration, N.~Aghanim {\em et~al.}, ``{Planck 2018
  results. VI. Cosmological parameters},''
  \href{http://dx.doi.org/10.1051/0004-6361/201833910}{{\em Astron. Astrophys.}
  {\bfseries 641} (2020) A6}, \href{http://arxiv.org/abs/1807.06209}{{\ttfamily
  arXiv:1807.06209 [astro-ph.CO]}}. [Erratum: Astron.Astrophys. 652, C4
  (2021)].

\bibitem{2019SJSC...41C.479B}
A.~H. {Barnett}, J.~{Magland}, and L.~{af Klinteberg}, ``{A Parallel Nonuniform
  Fast Fourier Transform Library Based on an ``Exponential of Semicircle''
  Kernel},'' \href{http://dx.doi.org/10.1137/18M120885X}{{\em SIAM Journal on
  Scientific Computing} {\bfseries 41} no.~5, (Jan., 2019) C479--C504},
  \href{http://arxiv.org/abs/1808.06736}{{\ttfamily arXiv:1808.06736
  [math.NA]}}.

\bibitem{2020arXiv200109405B}
A.~H. {Barnett}, ``{Aliasing error of the exp$(\beta \sqrt{1-z^2})$ kernel in
  the nonuniform fast Fourier transform},''
  \href{http://dx.doi.org/10.48550/arXiv.2001.09405}{{\em arXiv e-prints}
  (Jan., 2020) arXiv:2001.09405},
  \href{http://arxiv.org/abs/2001.09405}{{\ttfamily arXiv:2001.09405
  [math.NA]}}.

\bibitem{Zhang:2023oem}
Y.~Zhang, A.~R. Pullen, R.~S. Somerville, P.~C. Breysse, J.~C. Forbes, S.~Yang,
  {\em et~al.}, ``{Characterizing the Conditional Galaxy Property Distribution
  Using Gaussian Mixture Models},''
  \href{http://dx.doi.org/10.3847/1538-4357/accb90}{{\em Astrophys. J.}
  {\bfseries 950} no.~2, (2023) 159},
  \href{http://arxiv.org/abs/2302.11166}{{\ttfamily arXiv:2302.11166
  [astro-ph.GA]}}.

\bibitem{Mead:2020vgs}
A.~Mead, S.~Brieden, T.~Tr\"oster, and C.~Heymans, ``{HMcode-2020: Improved
  modelling of non-linear cosmological power spectra with baryonic feedback},''
  \href{http://arxiv.org/abs/2009.01858}{{\ttfamily arXiv:2009.01858
  [astro-ph.CO]}}.

\bibitem{COMAP:2021rny}
{\bfseries COMAP} Collaboration, D.~T. Chung {\em et~al.}, ``{A Model of
  Spectral Line Broadening in Signal Forecasts for Line-intensity Mapping
  Experiments},'' \href{http://dx.doi.org/10.3847/1538-4357/ac2a35}{{\em
  Astrophys. J.} {\bfseries 923} no.~2, (2021) 188},
  \href{http://arxiv.org/abs/2104.11171}{{\ttfamily arXiv:2104.11171
  [astro-ph.CO]}}.

\bibitem{Vernstrom:2013vva}
T.~Vernstrom, D.~Scott, J.~V. Wall, J.~J. Condon, W.~D. Cotton, E.~B. Fomalont,
  {\em et~al.}, ``{Deep 3 GHz number counts from a P(D) fluctuation
  analysis},'' \href{http://dx.doi.org/10.1093/mnras/stu470}{{\em Mon. Not.
  Roy. Astron. Soc.} {\bfseries 440} no.~3, (2014) 2791--2809},
  \href{http://arxiv.org/abs/1311.7451}{{\ttfamily arXiv:1311.7451
  [astro-ph.CO]}}.

\bibitem{COMAP:2021lae}
{\bfseries COMAP} Collaboration, D.~T. Chung {\em et~al.}, ``{COMAP Early
  Science. V. Constraints and Forecasts at z \ensuremath{\sim} 3},''
  \href{http://dx.doi.org/10.3847/1538-4357/ac63c7}{{\em Astrophys. J.}
  {\bfseries 933} no.~2, (2022) 186},
  \href{http://arxiv.org/abs/2111.05931}{{\ttfamily arXiv:2111.05931
  [astro-ph.CO]}}.

\bibitem{Behroozi:2019kql}
P.~Behroozi, R.~H. Wechsler, A.~P. Hearin, and C.~Conroy, ``{UniverseMachine:
  The correlation between galaxy growth and dark matter halo assembly from z =
  0\ensuremath{-}10},'' \href{http://dx.doi.org/10.1093/mnras/stz1182}{{\em
  Mon. Not. Roy. Astron. Soc.} {\bfseries 488} no.~3, (2019) 3143--3194},
  \href{http://arxiv.org/abs/1806.07893}{{\ttfamily arXiv:1806.07893}}.

\bibitem{Riechers:2018zjg}
D.~A. Riechers {\em et~al.}, ``{COLDz: Shape of the CO Luminosity Function at
  High Redshift and the Cold Gas History of the Universe},''
  \href{http://dx.doi.org/10.3847/1538-4357/aafc27}{{\em Astrophys. J.}
  {\bfseries 872} no.~1, (2019) 7},
  \href{http://arxiv.org/abs/1808.04371}{{\ttfamily arXiv:1808.04371
  [astro-ph.GA]}}.

\bibitem{Tinker:2008ff}
J.~L. Tinker, A.~V. Kravtsov, A.~Klypin, K.~Abazajian, M.~S. Warren, G.~Yepes,
  {\em et~al.}, ``{Toward a halo mass function for precision cosmology: The
  Limits of universality},'' \href{http://dx.doi.org/10.1086/591439}{{\em
  Astrophys. J.} {\bfseries 688} (2008) 709--728},
  \href{http://arxiv.org/abs/0803.2706}{{\ttfamily arXiv:0803.2706
  [astro-ph]}}.

\bibitem{Tinker:2010my}
J.~L. Tinker, B.~E. Robertson, A.~V. Kravtsov, A.~Klypin, M.~S. Warren,
  G.~Yepes, and S.~Gottlober, ``{The Large Scale Bias of Dark Matter Halos:
  Numerical Calibration and Model Tests},''
  \href{http://dx.doi.org/10.1088/0004-637X/724/2/878}{{\em Astrophys. J.}
  {\bfseries 724} (2010) 878--886},
  \href{http://arxiv.org/abs/1001.3162}{{\ttfamily arXiv:1001.3162
  [astro-ph.CO]}}.

\bibitem{Sato-Polito:2022wiq}
G.~Sato-Polito, N.~Kokron, and J.~L. Bernal, ``{A multi-tracer
  empirically-driven approach to line-intensity mapping lightcones},''
  \href{http://arxiv.org/abs/2212.08056}{{\ttfamily arXiv:2212.08056
  [astro-ph.CO]}}.

\bibitem{Niemeyer:2023yeu}
M.~L. Niemeyer, J.~L. Bernal, and E.~Komatsu, ``{SIMPLE: Simple Intensity Map
  Producer for Line Emission},''
  \href{http://arxiv.org/abs/2307.08475}{{\ttfamily arXiv:2307.08475
  [astro-ph.CO]}}.

\bibitem{Agrawal:2017khv}
A.~Agrawal, R.~Makiya, C.-T. Chiang, D.~Jeong, S.~Saito, and E.~Komatsu,
  ``{Generating Log-normal Mock Catalog of Galaxies in Redshift Space},''
  \href{http://dx.doi.org/10.1088/1475-7516/2017/10/003}{{\em JCAP} {\bfseries
  10} (2017) 003}, \href{http://arxiv.org/abs/1706.09195}{{\ttfamily
  arXiv:1706.09195 [astro-ph.CO]}}.

\bibitem{Sefusatti:2015aex}
E.~Sefusatti, M.~Crocce, R.~Scoccimarro, and H.~Couchman, ``{Accurate
  Estimators of Correlation Functions in Fourier Space},''
  \href{http://dx.doi.org/10.1093/mnras/stw1229}{{\em Mon. Not. Roy. Astron.
  Soc.} {\bfseries 460} no.~4, (2016) 3624--3636},
  \href{http://arxiv.org/abs/1512.07295}{{\ttfamily arXiv:1512.07295
  [astro-ph.CO]}}.

\bibitem{Blot:2018oxk}
L.~Blot {\em et~al.}, ``{Comparing approximate methods for mock catalogues and
  covariance matrices II: Power spectrum multipoles},''
  \href{http://dx.doi.org/10.1093/mnras/stz507}{{\em Mon. Not. Roy. Astron.
  Soc.} {\bfseries 485} no.~2, (2019) 2806--2824},
  \href{http://arxiv.org/abs/1806.09497}{{\ttfamily arXiv:1806.09497
  [astro-ph.CO]}}.

\bibitem{Hartlap:2006kj}
J.~Hartlap, P.~Simon, and P.~Schneider, ``{Why your model parameter confidences
  might be too optimistic: Unbiased estimation of the inverse covariance
  matrix},'' \href{http://dx.doi.org/10.1051/0004-6361:20066170}{{\em Astron.
  Astrophys.} {\bfseries 464} (2007) 399},
  \href{http://arxiv.org/abs/astro-ph/0608064}{{\ttfamily
  arXiv:astro-ph/0608064}}.

\bibitem{Bernal:2020pwq}
J.~L. Bernal, N.~Bellomo, A.~Raccanelli, and L.~Verde, ``{Beware of commonly
  used approximations. Part II. Estimating systematic biases in the best-fit
  parameters},'' \href{http://dx.doi.org/10.1088/1475-7516/2020/10/017}{{\em
  JCAP} {\bfseries 10} (2020) 017},
  \href{http://arxiv.org/abs/2005.09666}{{\ttfamily arXiv:2005.09666
  [astro-ph.CO]}}.

\bibitem{Bellomo:2020pnw}
N.~Bellomo, J.~L. Bernal, G.~Scelfo, A.~Raccanelli, and L.~Verde, ``{Beware of
  commonly used approximations. Part I. Errors in forecasts},''
  \href{http://dx.doi.org/10.1088/1475-7516/2020/10/016}{{\em JCAP} {\bfseries
  10} (2020) 016}, \href{http://arxiv.org/abs/2005.10384}{{\ttfamily
  arXiv:2005.10384 [astro-ph.CO]}}.

\bibitem{Barthelemy:2023mer}
A.~Barthelemy, A.~Halder, Z.~Gong, and C.~Uhlemann, ``{Making the leap I:
  Modelling the reconstructed lensing convergence PDF from cosmic shear with
  survey masks and systematics},''
  \href{http://arxiv.org/abs/2307.09468}{{\ttfamily arXiv:2307.09468
  [astro-ph.CO]}}.

\end{thebibliography}\endgroup
\bibliographystyle{utcaps}

\end{document}